\documentclass[10pt,journal,compsoc]{IEEEtran}
\usepackage[nocompress]{cite}
\usepackage[pdftex]{graphicx}
\usepackage{amsmath}
\usepackage{array}\usepackage{subfigure}
\usepackage{mathtools}
\usepackage{stfloats}
\usepackage{supertabular}
\usepackage{url}

\usepackage[breaklinks,pagebackref=false,colorlinks,bookmarks=false]{hyperref}
\hypersetup{
     colorlinks=true,
     linkcolor=black,
     citecolor = black,      
     urlcolor=black,
     }
\usepackage{footmisc}
\usepackage{caption}
\usepackage{multirow}
\usepackage{multicol}
\usepackage{makecell}
\usepackage{paralist}
\usepackage{enumerate}
\usepackage{enumitem}

\usepackage[switch]{lineno}
\usepackage{comment}

\begin{document}
\bstctlcite{IEEEexample:BSTcontrol}
\title{Iris Recognition Performance in Children: A Longitudinal Study }

\author{Priyanka~Das, Laura~Holsopple, Dan~Rissacher, Michael~Schuckers and~Stephanie~Schuckers
        
\IEEEcompsocitemizethanks{\IEEEcompsocthanksitem P. Das, L. Holsopple and S. Schuckers ar with the Department
of Electrical and Computer Engineering, Clarkson University, Potsdam,
NY, 13699.\protect\\
E-mail: prdas@clarkson.edu
\IEEEcompsocthanksitem D. Rissacher is presently an independent researcher at his own company CRIA Corp.
\IEEEcompsocthanksitem M. Schuckers is with the St. Lawrence University.}}
\IEEEtitleabstractindextext{%
\begin{abstract}
 There is uncertainty around the effect of aging of children on biometric characteristics impacting applications relying on biometric recognition, particularly as the time between enrollment and query increases. Though there have been studies of such effects for iris recognition in adults, there have been few studies evaluating impact in children. This paper presents longitudinal analysis from 209 subjects aged 4 to 11 years at enrollment and six additional sessions over a period of  3 years. The influence of time, dilation and enrollment age on iris recognition have been analyzed and their statistical importance has been evaluated. A minor aging effect is noted which is statistically significant, but practically insignificant and is comparatively less important than other variability factors.  Practical biometric applications of iris recognition in children are feasible for a time frame of at least 3 years between samples, for ages 4 to 11 years, even in presence of aging, though we note practical difficulties in enrolling young children with cameras not designed for the purpose. To the best of our knowledge, the database used in this study is the only dataset of longitudinal iris images from children for this age group and time period that is available for research. 
\end{abstract}

}
\maketitle

\IEEEraisesectionheading{\section{Introduction}\label{introduction}}
\IEEEPARstart{D}{igital} identity and the need for convenient secure authentication have contributed to the proliferation of automated biometric recognition in the government and consumer space from the late 1990s until today. Applications include government sector (border security, criminal investigation, national security, citizenship registration, benefit distribution, human trafficking) and commercial sector (security, banking, personal identification, healthcare). \textit{`Persistence'} (permanence) and \textit{`distinctiveness} (individuality) \cite{jain2015guidelines} inherent to individuals are core characteristics of biometric recognition.  \par

The possible individuality and permanence of the iris led the ophthalmologist Adler in 1965 to consider it for biometric recognition - \textit{``The markings on the iris are so distinctive that it has been proposed to use photographs of iris as a means of identification instead of fingerprints"} \cite{adler1965physiology}. It is a general assumption that the iris is \textit{`highly stable'} \cite{ma2004efficient} \cite{daugman1994biometric} \cite{daugman1993high} \cite{wildes1997iris} \cite{flom1987iris} throughout a person's lifetime. However, datasets supporting this hypothesis are limited, challenging to obtain and largely limited to adults. As children grow and mature, one might conjecture that biometric traits, including the iris, might change. Hence, results for adults may not apply to children. If we wish to use iris biometric recognition in applications involving children (e.g. keeping vaccination records in developing countries, investigation of human trafficking), it is important to determine how well iris recognition works for children. \par

The present pressing questions from a technical and practical viewpoint are - 
\begin{itemize}[leftmargin= *, noitemsep]
    \item \textit{\textbf{Does aging change the iris structure to a point that it impacts the use of iris for biometric recognition in children?} }
    \item \textit{\textbf{If growth in children impacts iris recognition performance, is there an age at which these impacts are no longer seen?}}
\end{itemize}
This paper works towards answering issues related to iris aging in children through analyzing the iris of the same children collected every 6 months, for over 3 years from 209 children in the age group of 4 to 11 years at enrollment.\par 

International Organization for Standardization (ISO) defines biometric \textbf{`reference aging'} as \textit{``change in error rates with respect to fixed reference caused by time-related changes in the biometric characteristics"} \cite{iris_standard_report}. The National Institute of Standard and Technology (NIST) defines iris aging as \textit{``irreversible changes to the anatomy, primarily the iris texture"} \cite{grother2013irex}. In this work we investigate whether iris recognition accuracy decreases with the time lapsed between collection of initial enrollment and subsequent recognition images in children. Our analysis includes investigating the match scores for false match and false non-match errors, identifying the causes of the errors, and investigating the factors contributing to match score variation like enrollment age, dilation, and dilation constancy in addition to aging. A linear mixed effects model has been designed and used to understand the effect of different factors on match score variability.\par 

\subsection{State of Art}\label{SOA}
Though the idea that the iris could be used to recognize people goes back to at least Bertillon \cite{bertillon1896signaletic}, modern automated, iris recognition began in 1987 with a patent by Flom and Safir\cite{flom1987us}, followed by Daugman's development of the algorithm of iris pattern coding for automated visual recognition \cite{daugman1993high}\cite{daugman1994biometric}. Most research has concentrated on the technical development and improvement of iris recognition systems. The effect of aging on iris for biometric recognition has been a minimally studied area with most researchers initially echoing the assumption of high temporal stability of iris with little evidence to support this claim. Studies began emerging from 2008. 

Between 2008 and 2014 multiple research groups investigated the temporal stability of iris as a factor of aging \cite{tome2008effects} \cite{baker2009empirical} \cite{baker2013template} \cite{fenker2012analysis} \cite{rankin2012iris} \cite{sazonova2012study} \cite{czajka2013template} and noted increased False Non-Match Rates (FNMR), increased Hamming distance between images or decay in genuine match scores (MS) in longitudinal scenario. Researches in \cite{sazonova2012study} and \cite{czajka2013template} drew notice to the importance of quality factors in iris recognition and found a lack of evidence to conclude aging of iris texture as pupil dilation, iris diameter and occlusion of iris can impact performance.\par
In the wake of these studies, in 2013 NIST re-evaluated the studies by Baker \cite{baker2009empirical} and Bower \cite{baker2013template} and published their report - IREX VI, \cite{grother2013irex} which concluded that the increased FNMR in \cite{baker2009empirical} \cite{baker2013template} was the result of variation in dilation over the collection period. The report also analyzed and modelled a large match score log from a border crossing application (NEXUS) and concluded minimal ageing over a 10 year time-span, much smaller than typical day to day variations. These results were primarily from adults. NIST was unaware of a small portion of the children population enrolled in NEXUS \cite{gorodnichy2018analysis} and did not take into account in their analysis \cite{James}. Change in dilation is excluded from consideration by their definition of aging as dilation varies stochastically on a \textit{"timescale ranging from below one second up to several decades''}, impacted by factors including environmental factors or disease and it can be mitigated by external illumination and other hardware or software solutions. This motivated discussion about the very definition of \textbf{`iris aging'}, approaches for statistical regression modelling characterizing relationship between variables \cite{ND_response_to_NIST} \cite{model_paper_nist} and impact of retrospective dataset structure on modelling. The NEXUS dataset was influenced by 'truncation' and 'censoring' \cite{ortiz2015exploratory} which might impact the modelled estimate of aging. The difficulties around usage of retrospective large operational datasets for studying age related impacts on biometric recognition, specifically iris, was explored and analyzed in \cite{ortiz2016pitfalls}. Research results may be impacted due to heterogeneity in the datasets because of data collected in multiple locations, different times, uncontrolled environment, seasonal impact and impacted by variability in lighting. With the exception noted, the prior works described above were specific to adults. Our own research attempts to perform similar analyses on children. Prior work related to children is described later in this section.    \par

Various research has found that pupil size/dilation varies as a function of age\cite{birren1950age_PD} \cite{adler1965physiology} \cite{winn1994age_PD} \cite{ortiz2013MS_PD} \cite{Das_dilaiton} where pupil size is small for newborns until the first year of life, reaching its maximum size in childhood and adolescence and then gradually becoming smaller with advancing age \cite{adler1965physiology} and the decay is linear (age: 17 and above ) with increasing age at different luminance levels\cite{winn1994age_PD}. A measurable degradation in MS due to dilation differences at different ages has been noted in adults with a corresponding increased FNMR \cite{ortiz2013MS_PD} by applying a linear regression model. In the statistical analysis for our paper, we consider the effect of dilation on performance and factors including age and environment.\par
Research on aging effects on biometric iris recognition performance in children has been limited primarily due to the scarcity of longitudinal iris data on children. Three factors that contribute to this scarcity are - (1) the recent introduction of iris recognition for scenarios including children; (2) the Institutional Review Board (IRB) regulations for children \cite{IRB_protocol}; and (3)  the lack of iris capture systems specifically designed for children. In 2017 Basak et al.  \cite{basak2017multimodal} created the first public multimodal dataset \cite{CMBDDatabase}, including iris modality, from approximately 100 children aged between 18 months and four years, and evaluated the feasibility of data collection from toddlers and pre-school children with the equipment available in the market. The study concluded (i) though iris data capture is challenging in younger children, it yields best performance - 99.82\% single iris accuracy (98.95 left) for iris verification and (ii) 100\% recognition accuracy with multiple iris images. In 2019 Nelufule et al. worked on iris quality assessment algorithm, using a proprietary dataset of 103 subjects in the age group of six weeks to five years \cite{nelufule2019image}, to identify and eliminate noise data like occluded iris, light variation, off angle, pupil dilation. No information is provided on the availability of the database. The assessment concluded that child data produces similar quality distribution as that of an adult if the unusable images are removed from the child dataset. \par

 In real-world applications, the Aadhaar program in India, which creates a unique ID using fingerprint and iris for its population, limited enrollment age at 5yrs with re-enrollment at the age of 15yrs, and updating the biometric data every 10yrs \cite{Andhar}. No supporting reason has been offered by UIDAI for the age limits. Report on the Nexus program by the Canadian Defence Research Organization \cite{Canada_Report}, which uses iris biometric recognition with no age limit, mentions a child was enrolled at an age of 8 months and verified successfully at the age of 9 months, 12 months and 14 months. Based on a study of the transaction logs, the report pronounced - (1) the size of the iris changes for children is not stable until 6 yrs to 8yrs of age; (2) enrollment of both eyes is much more difficult for children under 14. The subjects below 14 had a higher percentage of ‘enrollment with only one eye'; (3) successful enrollment in the younger age group (under 14 years) are substantially low compared to the middle age group, which can be attributed to less frequent younger age travellers using the system. However, no information is provided for unsuccessful enrollments in any age group. The report concludes that iris recognition is less reliable but still useful for younger age groups.\par

In 2018, our group issued a preliminary report focused on iris recognition in children \cite{johnson2018longitudinal} with three collections spaced by 6 months over one year - a subset of the dataset that we use in this paper. The paper reported on inter-session match score variation and age group analysis and concluded biometric characteristic stability over a period of one year from 123 children aged four and above. As a continuation of this work \cite{johnson2018longitudinal}, we extended the analysis of the same data, now seven collections over three years, to study the longitudinal effect of aging in iris recognition of children. Studies before Johnson et. al. \cite{johnson2018longitudinal} in 2018 were not longitudinal, i.e., the same subjects were not followed up for more than six months, or the subjects that were followed were adults. The focus age group of our research is 4 to 14 years. In addition, many of the prior studies concentrated on FNMR performance and did not explore the root causes for the match score effects that drive the FNMR. We believe that understanding root causes is essential in making reliable decisions on temporal stability of biometric recognition systems. To the best of our knowledge this is one of the first longitudinal studies exploring the aging effect on iris as a biometric in children.\par 
The rest of this paper is organized in the following order: Section~\ref{Methodology} details the dataset, collection protocol, data statistics and provides an outline of the data processing. Section~\ref{LMER} details the Linear Mixed Effects (LME) model that we use in our analysis. Section~\ref{sec:DC} provides a discussion on overall image quality and dilation constancy. Section~\ref{Results} reports our results. Section~\ref{Limit} discusses the limitations of our analysis and suggests possible future work. Section~\ref{conclusion} summarizes answers to the questions posted in Section~\ref{introduction}.\par

\section{Methodology} \label{Methodology}

The research in this paper is part of a larger longitudinal study of biometric characteristics in children. The age of participants in the study spans 4yrs to 14yrs. The dataset consists of six modalities including iris. The research team collaborates with the local elementary and middle school to identify and enroll subjects for voluntary participation, in accordance with an approved IRB protocol.\par
 The collection equipment is set up in an isolated room provided by the school for the entire duration of the collection week. However, the rooms may vary with availability at each session. Thus it is expected that there may be changes in the collection environment due to factors such as lighting and noise which may impact biometric measurements. Measures to mitigate the impact of environmental factors are taken, including drawing the blinds in the room to prevent exposure from external daylight and providing participants some leisure time to get accommodated to the room environment like lighting, body temperature, and humidity. The protocol allows participants voluntary participation on a given day of collection with the option to abstain from participating; however, any personal emotional state remains unaccounted. The same equipment, Iris Guard IG-AD100 Dual Iris Camera, has been used for all collections. \par
 Starting in January 2016, the study covers three years, with an approximate time interval of 6 months, with seven sessions included as of November 2019. 239 subjects were enrolled; however, not all subjects participated in all sessions. The variation in the number of participants for each session is impacted by newly enrolled participants each school year, absentees, those unwilling to participate on a given day, and participants whose families have moved out of the school district. 209 subjects have participated in more than one session. Enrollment age varies from 4yrs to 11 yrs, with one subject with an enrollment age of 3yrs. After the first year, enrollment was offered to only pre-kindergarten class, which are primarily 4 years old. To maintain privacy, we use only the birth-year of the participants. Thus, the ages are approximate. The age of three subjects were approximated based on their grade at enrollment as their birth-year was unavailable. The enrollment age breakdown is shown in Table~\ref{table:EA}.

\begin{table}[!t]
\scriptsize
\centering
\caption{\textbf{Enrollment Age Statistics}}
\label{table:EA}
\begin{tabular}{|c|c|c|c|c|c|c|c|c|c|c|}
\hline
\textbf{Age} & 4 & 5 & 6 &7 & 8 & 9 & 10 & 11 & NA  \\
\hline
\textbf{Count} & 39 & 20 & 31 & 28 & 26 & 24 & 29 & 10 & 2\\
\hline

\end{tabular}
\end{table}

\begin{figure}[htbp]
    \centering
    \subfigure{(a)\label{fig:Count_image_subject}\includegraphics[width=4.3cm,height=4.2cm]{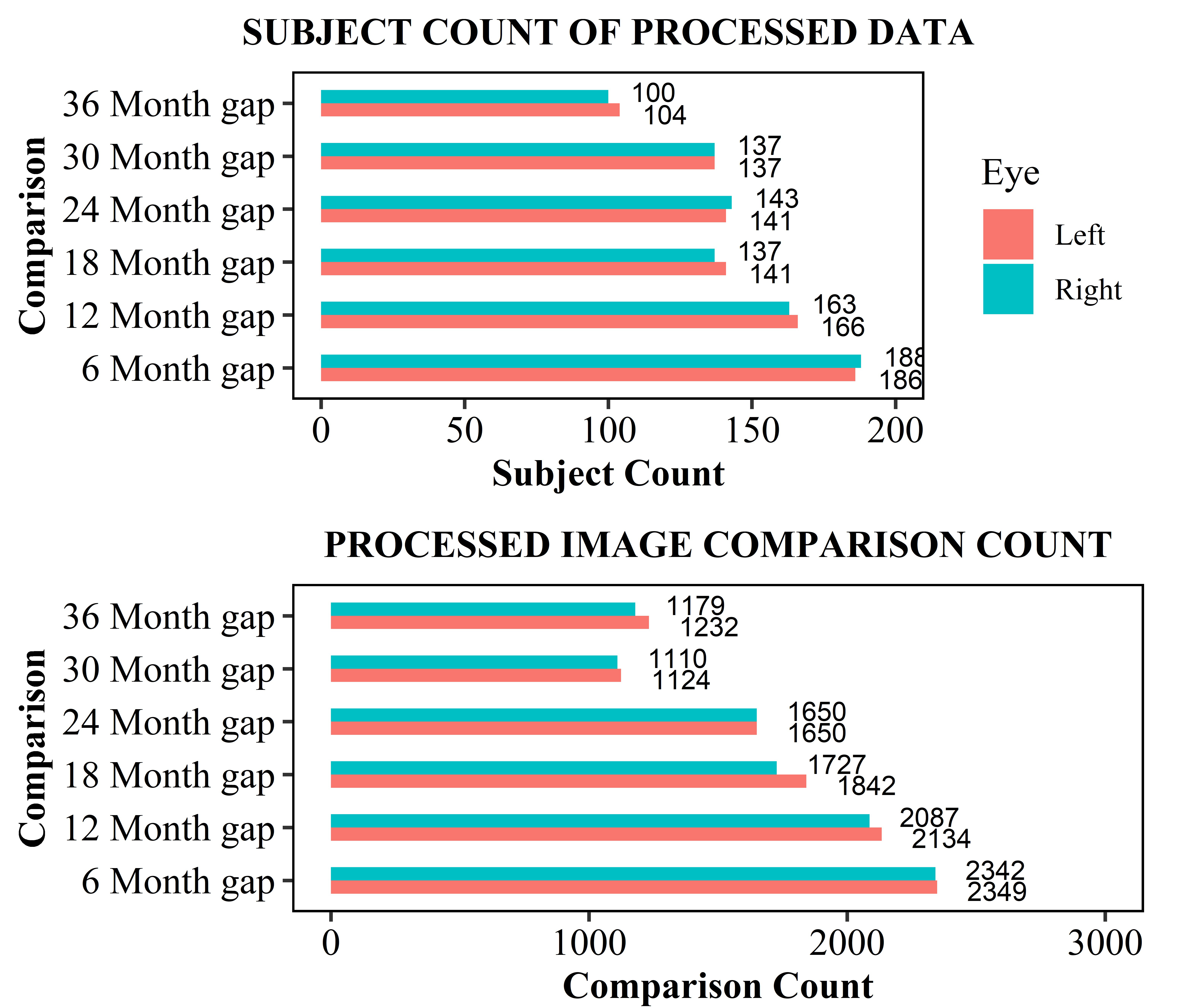}}
    ~
    \subfigure{(b)\label{fig:Arm}\includegraphics[width=3.2cm,height=4.2cm]{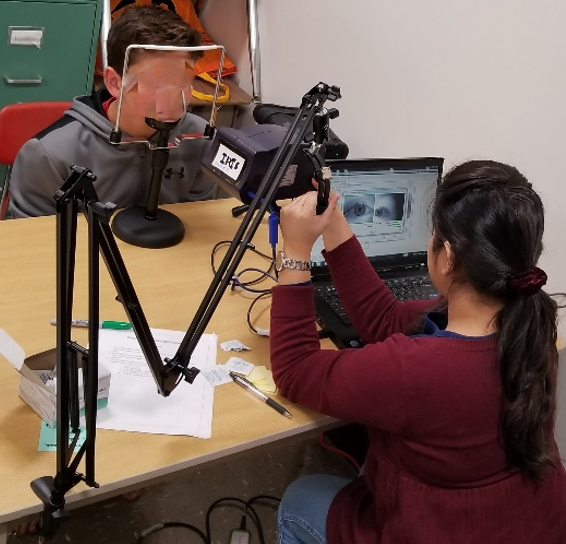}}
  
    \caption{\footnotesize (a) Summary of subject count (top and comparison count (bottom) for right and left iris   by time between sessions(6,12,18,24,30 and 36 months) in this study; (b) An articulating arm mount for the Iris ID camera }
\end{figure}

\subsection{Iris Data Collection} \label{iris_coll}
The Iris Guard IG-AD100 Dual Iris Camera \cite{IrisGuard} provides ISO/IEC 29794-6 \cite{iris_standard_report} compliant iris images in the NIR wavelength. The sensor has auto-focus and a mirror which assists participants to self-adjust their position relative to the sensor. Once the subjects are correctly positioned, both the eyes are captured, with a delay of a few seconds between the left and right iris capture. In addition to the NIR illuminators, the sensor also uses a flashing white light intended to stabilize the subject's pupil dilation. Capturing the iris was more difficult in younger subjects (4 - 5 yrs), as they have difficulty holding their eyes still, opening their eyes wide enough for the sensor to sense and capture the data, and not tilting their head relative to the camera. Participants were asked to place their chin on an eye-examination grade chin rest for stability, and collectors adjusted the height, distance and positioning of the camera. Some subjects still had difficulty with iris capture resulting in poor quality images as well as failure to capture. We included an articulating mount for the camera, shown in Figure~\ref{fig:Arm}, from the 6th session to provide more flexible control on part of the collector. With this, the collector had control of the position of the camera instead of asking the subjects to adjust their position with respect to the camera. This reduced the issue of failure to capture by a large degree. A second commercially available iris sensor, iCAM T10 by Iris ID, was introduced from the 4th Collection due to its binocular form factor. However, this paper analyzes the data collected only from IG-AD100.

 \subsection{Failure to Acquire (FTA) \label{FTA}}
 Data from approximately 25 subjects were not acquired at one or more sessions, even though they participated for other modalities. This includes failure to acquire (FTA), refusal to participate, health issues or time-constraint for collection. More details are provided in the Discussion section and this is the focus of future work.
 
\subsection{Iris Matcher}\label{Verieye}
 A commercially available SDK, VeriEye \cite{VerieyeSDK}, by Neurotechnology, was used for image comparison and deriving all attributes used in this study. The iris recognition software and all metrics considered for analyses follows ISO/IEC 29794-6 \cite{iris_standard_report}. Based on the IREX X \cite{ngan2019irex} report, Neurotechnology is one of the top two contenders in every field of analysis. Matching was done on VeriEye version 10.0 and all other attributes were obtained from VeriEye version 11.1. VeriEye computes a similarity score or match score (MS) in the range of 0 and 1556 where a higher score indicates greater similarity; it also computes the pupil and iris radii, expressed in pixel units. VeriEye provides a threshold calibration of match score vs. FAR; for an FAR of 0.1\% the corresponding match score threshold is 36\cite{VerieyeSDK}. All analysis has been done at 0.1\% False Match Rate (FMR) as suggested in IREX VI \cite{iris_standard_report}. Additional analyses were performed in MATLAB R2018b and RStudio.
 
\subsection{Statistics of the Data}\label{sec:Data_stat}
418 irides from 209 subjects were processed. There is a variation in the capture between right iris (RI) and left iris (LI), due to enrollment failure.  \par
Two images were captured for each eye at the first and sixth session; for all other sessions, four images were captured per eye. Images collected before the seventh session had high intra-session correlation due to the internal setup of the camera which captures multiple images within seconds. The collection protocol was modified in the seventh session. Four images were captured per eye in two different sets spaced by a time gap of approximately two minutes. Participation count of  number of subjects and the number of comparisons for different time-frames are summarized in Figure~\ref{fig:Count_image_subject}. \par
\begin{figure}[!t]
\centering
    \includegraphics[scale=0.7]{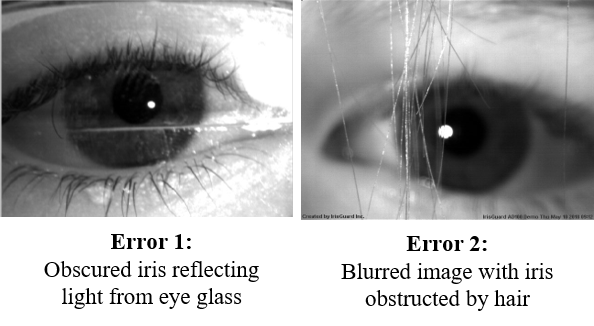}  
    \caption{ \footnotesize Examples of images of two subjects removed from the dataset on manual cleaning. Obvious errors were identified (obscured iris and blurred iris). A total of six images were removed. Two images had Error 1 and four images had Error 2. }
    \label{fig:Error}
 \end{figure}

Six images from two different subjects with clearly observable artifacts were discarded from the database based on a manual cleaning. Two different errors were identified in those images - four images were blurred with iris obstructed by hair and two images had an obscured iris with reflecting light from eye glasses. Examples are shown in Figure~\ref{fig:Error}.\par 
The first session that a subject participated is considered as the enrollment session. Images collected at the enrollment session are mated with all images from subsequent sessions. Intra-session image comparison is not considered in this study due to:
\begin{itemize}[leftmargin= *, noitemsep]
\item intra-session image comparison does not contribute to a real world application scenario
\item the measures have no contribution towards how increased time between enrollment and probe impact performance.
\end{itemize}
The data has been sectioned and analyzed in three ways: G1: analysis of all 209 subjects who participated in more than one session; G2: analysis of a subgroup of 105 (RI:101) subjects who participated in at least 1st and 7th session with intermittent absentees; and G3: analysis of a subgroup of 63 (RI:62) subjects who participated in all seven sessions. LI and RI irides were processed separately. The total number of comparisons per subject are non-uniform. A total of 20428 (right: 10096, left: 10332) comparisons were performed; 6421 (LI: 3216; RI: 3205) probe images were compared against 1301 (LI: 658; RI: 643) enrollment images from 209 subjects.

\section{Linear Mixed Effects Model}\label{LMER}
Ordinary Least Square (OLS) regression modelling gives exploratory insight into the match score (MS) as time increases between enrollment and query. A linear functional form is often used for individual growth modelling  based on visual inspection of the individual patterns, constricted time-duration of the data (3 years) and referencing the literature \cite{singer2003applied} \cite{grother2013irex}.
However, modelling longitudinal data of sampled data sets for constricted time duration to explore changes in a population is a challenge due to the following reasons:
\begin{itemize} [leftmargin=*,noitemsep]
 \item Different subjects will have a different trajectory of the empirical growth;
 \item A simple linear model might not be a proper representation of accumulative intra-individual and inter-individual variation for longitudinal data; and
 \item Variation in MS may reflect other factors in addition to ageing like random fluctuations of different factors (error or quantitatively unaccountable factors like environment, medication, illumination, emotion etc).
\end{itemize}

Linear Mixed Effects Models address the disparity between within-individual variation, and inter-individual differences. \textit{``Mixed Effects"} refers to a combination of \textit{fixed effects} and \textit{random effects} on the response variable. \textit{Fixed effects} quantify the response variable with respect to the effects of predictors on the inter-subject variation and \textit{random effects} quantify any possible variation in the response that could be accounted for by predictors on the intra-subject variation.  The basic model structure is defined as:
\begin{equation}
\footnotesize
\begin{split}
 \mathit{\mathbf{Response \sim F\_expression +}}  \mathit{\mathbf{(R\_expression| Factor )}}
\end{split}
\end{equation}
Here, \textit{F\_expression} represents the fixed effects model matrix. Each random effect is represented by: \textit{$(R\_expression | Factor )$} 
where, \textit{R\_expression} is the random effect predictor having different effects on the \textit{Response} variable for each level of grouping factor, i.e. \textit{Factor}.

All modelling and analysis has been computed on the R platform using the package \textbf{`lme4'} \cite{lme4}. When a restricted portion of lifespan is being analyzed, as in our case, the model does not predict or conclude anything beyond the time-frame of research.

\subsection{LMER Model and Predictors}\label{model_pred}
The variation in MS is modelled using the predictors - time difference between collection of enrollment and probe image (TD), enrollment age (EA), probe dilation (PD) and dilation difference between probe and enrollment image ($\Delta$D) as described below.

\begin{enumerate}
    \item \textbf{Time Difference (TD)}
    The difference in time between the enrollment and probe image measured in \textit{``months"}.
    
    \item \textbf{Enrollment Age (EA)}
    Age when the subject is enrolled in the study, i.e. the age of their first collection in \textit{``years"}. 

    \item \textbf{Dilation (D)}\label{Dil}
    Dilation or pupil dilation is a dimensionless quantity measuring the degree to which the pupil is dilated or constricted, measured as a ratio of pupil radius and iris radius. The measure follows ISO/IEC 29794-6 \cite{iris_standard_report} as defined below.
        \begin{equation}\label{eq:D}
        \footnotesize
         \mathit{\mathbf{Dilation(D) = \frac{Pupil\: radius}{Iris\: radius} \times 100}}
        \end{equation}
    The dilation of the probe image is considered for the model denoted by PD.

    \item \textbf{Delta Dilation ($\Delta D$)}
    Difference in the pupil dilation between a mated pair of iris images. The measure follows NIST work in \cite{grother2013irex} as below.
        \begin{equation}
        \footnotesize
           \mathit{\mathbf{ Delta\:Dilation (\Delta D) = 1 - \frac{1-\frac{D1}{100}}{1-\frac{D2}{100}}}}
        \end{equation}
    considering, D1 $\geq$ D2, where, D1 and D2 are the pupil dilation of the first and the second iris images as estimated by equation \ref{eq:D}.
\end{enumerate}

The designed model to predict the match score is represented in equation~\ref{eq:Full Model}. The model considers linear change in the match score for up to 36 months between enrollment and query. Though the modelling of the MS is considered linear statistically, non-linear changes in the predictors are taken into account by including second order terms. To the best of our knowledge, there is no prior indication or evidence of non-linear change in the iris. We have tested different configurations of the model. The model in Equation~\ref{eq:Full Model} was concluded to be the best fit. This model has been used to interpret results and is extensively discussed in section~\ref{LMER_model_analysis}.

\begin{equation}
\footnotesize
\label{eq:Full Model}
    \begin{split}
    \mathit{\mathbf{MS \sim  \boldsymbol{\beta}_{0} + \boldsymbol{\beta}_{1}TD  +\boldsymbol{\beta}_{2} EA + \boldsymbol{\beta}_{3} PD + \boldsymbol{\beta}_{4} \Delta D }}  \\
     \mathit{\mathbf{ +  \beta_{5} EA^2 +\beta_{6} \Delta PD^2 + \boldsymbol{\beta}_{7} \Delta D^2  + b_{0i} }} \\
      \mathit{\mathbf{ +  b_{1i}TD  + b_{3i}PD + b_{4i} \Delta D + b_{6i} PD^2 }} 
    \end{split}
\end{equation}

 where,  \\
 \begin{itemize}[leftmargin=*,noitemsep]
  \item $\beta_{k}$ is the fixed regression coefficient for corresponding parameter, \textit{k}.
   \item $b_{ki}$ is the random regression coefficient for corresponding parameter, \textit{k}, for subject, \textit{i}.
  
   \item $\beta_{0}+ b_{0i}$ is the sum of fixed and subject specific random intercept corresponding to the initial state.
   
   \item $\beta_{k}+ b_{ki}$ is the subject specifi(\textit{i}) gradient for the corresponding parameter \textit{k}.
   \item Time Difference (TD) between enrollment and subsequent verification has been considered for both fixed and random effects to account for possible variability  in MS due to both intra-subjects and inter-subject effects of TD. 
   \item Enrollment Age (EA) has been considered only for fixed effect. All other predictors are considered for both fixed and random effect. 
   \item Both linear and quadratic factors are introduced for variables: dilation difference ($\Delta D$) , probe dilation ( PD) and enrollment age (EA) for fixed effects. We assume the non-linearity in modelling with respect to match score (MS) for inter-subject data.
   
   \item Both linear and quadratic factors are considered for probe dilation in random effects for intra-subject data, to account for any non-linearity in dilation for each subject over time.
   
   \item We consider correlation between all coefficients associated with the same random effect term i.e. for each subject we consider correlation between coefficients of TD, PD, $\Delta D$ and $ PD^2 $.
\end{itemize}

\section{Overall Image Quality and Dilation Constancy (DC)}\label{sec:DC}
\begin{figure}[!t]
\centering
    \includegraphics[height= 1.5in, width= 3.5in]{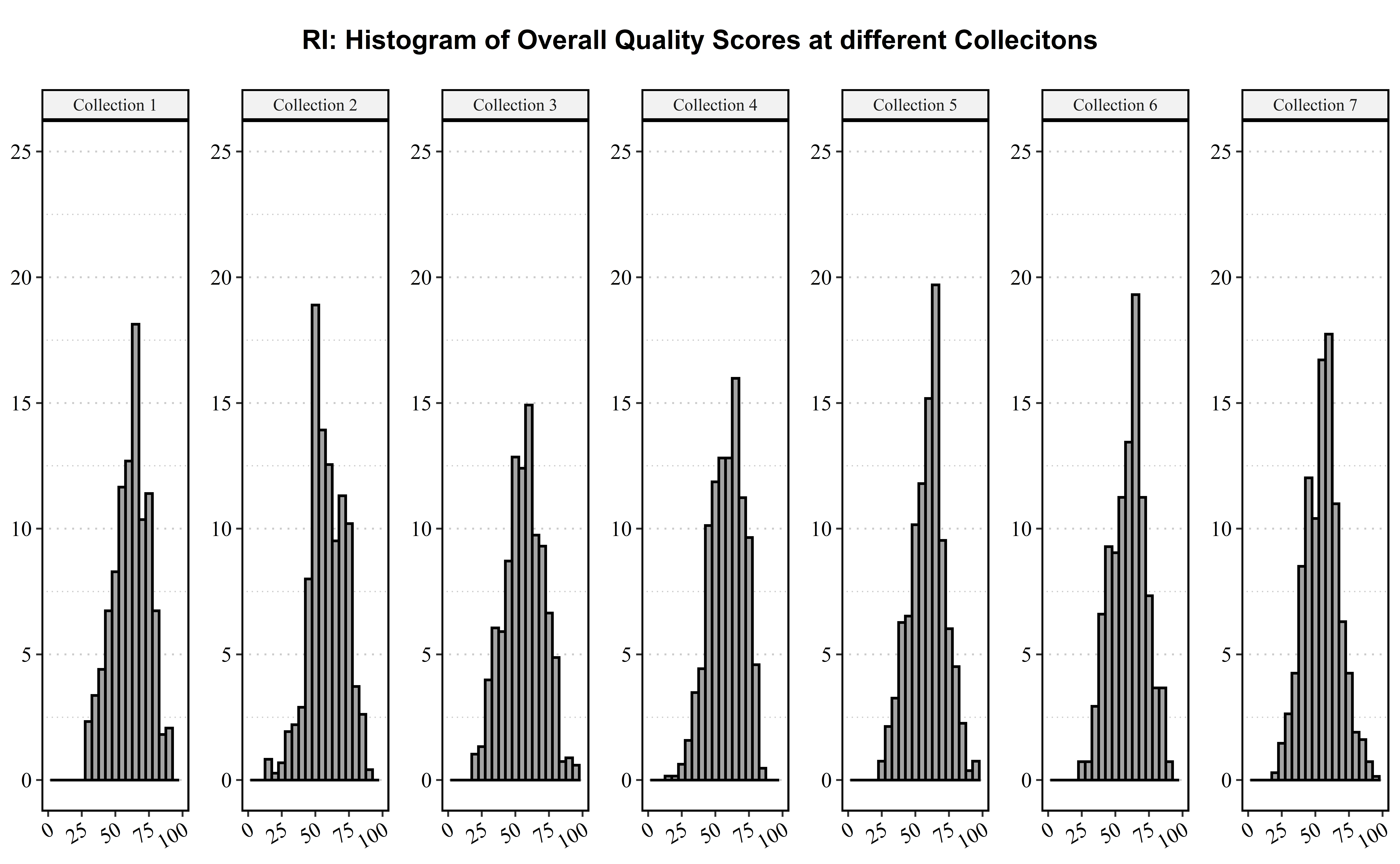}  
    \caption{\footnotesize Distribution of overall quality score of all images from Collection 1 through Collection 7 (left to right)}
    \label{fig:overall_quality}
 \end{figure}

ISO/IEC 29794-2 \cite{iris_standard_report} defines quality measures for a single image, which can be combined to a single scalar overall iris quality score ranging from 0 to 100; higher value is better quality. The score reflects the expected performance of the iris image. The overall image quality scores of all images collected in the seven sessions are derived from VeriEye\cite{VerieyeSDK} and the distribution is shown in Figure \ref{fig:overall_quality}. The t-test statistics show there is a statistically significant (p $<$ 0.001) difference in the mean of the overall quality scores between collection sessions when compared to Collection 1 except for Collection 6 (p$>$0.5). However, as the mean values are - 57.79, 51.72, 54.53, 55.97, 54.95, 55.97, 54.95 for Collection 1 through  Collection 7, the difference in the mean quality scores are practically not significant. Pearson's correlation between image quality score and age at collection of the image spanning 4 to 14 years is weak with correlation coefficient of 0.14. \par 
Dilation Constancy (DC) evaluates the similarity in pupil dilation in a mated pair of iris images. It is a dimensionless quantity measured as below.
\begin{equation}
\small
    \mathit{\mathbf{Dilation\:Constancy(DC) = \frac{100 - max(D1,D2)}{100 - min(D1,D2)}}}
\end{equation}
where D1 and D2 are the pupil dilation of the first and the second iris images as estimated in equation \ref{eq:D} as detailed in Section \ref{Dil}. Delta Dilation depends on the variability of both D1 and D2. If either of D1 or D2 changes, delta dilation changes. If the delta dilation increases, the dilation constancy decreases and vice-versa. DC has been used to understand the match score variability with DC and any relation between false non-match and DC. DC and Delta Dilation are related: $\Delta$D = 1 - DC

\section{Results \& Discussion}\label{Results}
This section reports the analysis of the performance of biometric recognition of iris in children with increasing time difference between enrollment and probe in presence of variable factors - enrollment age, dilation, and dilation difference between enrollment and probe.  The next sections cover modelling of the match scores using linear mixed effect model for groups based on participation and age, dilation and dilation constancy, false non-match rate, detailed examination of the false non-match cases and ROC of the dataset. The overall goal is to assess the viability of biometric applications of iris recognition in children as they age. 

\subsection{Iris Match Score Analysis: Participation based Groups}\label{P_Group}

\begin{figure}[!t]
\centering
    \includegraphics[width=3.2in, height=1.8 in]{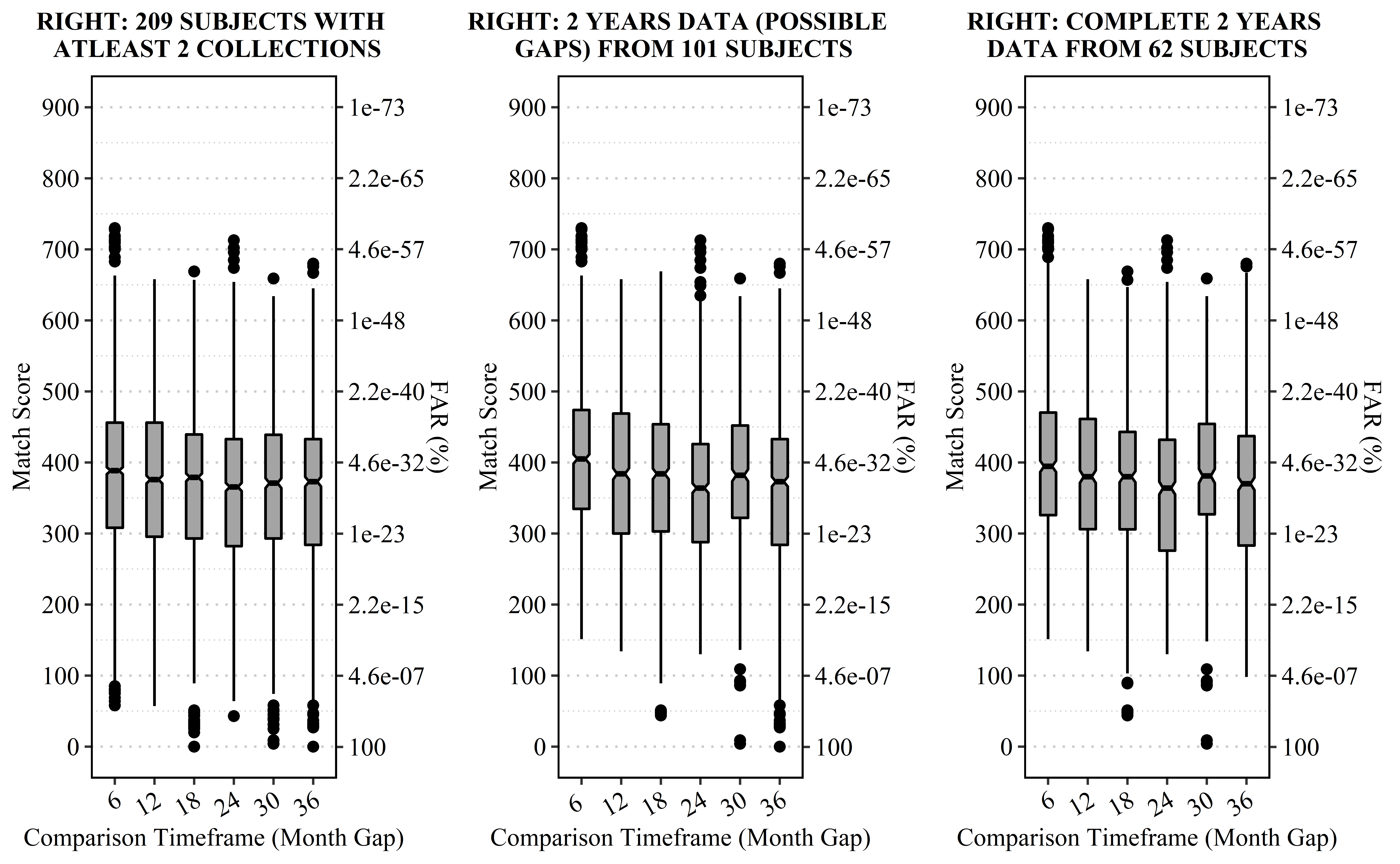}  
    \caption{\footnotesize Box-plots of all scores from individual subjects for right iris grouped based on participation: G1 (left) : all subjects have at least one mated pair data; G2 (middle): all subjects have mated pairs from at least at 6 and 36 month TF; G3 (right): all subjects have mated pairs from all TFs (6, 12, 18, 24, 30, 36 months). Relative False Accept Rate (FAR) w.r.t the MS is shown in the right y-axis. }
    \label{fig:group_boxplot}
 \end{figure}

Based on participation we have three groups: G1, G2 and G3 (refer Section~\ref{sec:Data_stat}). We have analyzed MS following the model described in Section~\ref{model_pred}. A visual summary of the data is shown in Figure~\ref{fig:group_boxplot}. Extensive analysis of the results from the model for G1-LI addressing the impact of aging on iris recognition performance in children are detailed in the following section. The same conclusions are true for all three groups with only a small variation in coefficients. 

 \subsubsection{Iris Match Score Modelling}\label{LMER_model_analysis}
The Linear Mixed Effects Regression Model (LMER) addresses the questions - \textit{ Are the explanatory/predictor variables - time difference (TD), enrollment age (EA), pupil dilation (PD), change in dilation between mating pair of images ($\Delta D$) a significant predictor of the response variable (match score)? Is there a significant decrease in the iris match scores in children as time between enrollment and query increases?} The LMER model with random slope for each subject is designed as shown in Equation~\ref{eq:Full Model} and detailed in Section~\ref{LMER}. We have referenced the model used by NIST in IREX VI \cite{grother2013irex} as a base model and augmented it with additional terms like enrollment age as a quadratic factor. Enrollment age is added to account for the variable age of the subjects in our dataset. The right and left iris are modelled separately. The multi-collinearity between independent variables are eliminated by re-centering the following variables - PD, EA and $\Delta D$, following discussion in \cite{ND_response_to_NIST}. The fixed effect of the designed model considers the overall inter-subject variability of TD, EA, PD, $\Delta D$  and impact of these variabilities on the MS of the entire dataset representing the population. Approximately 5.54\% (4.03\% for RI) variation in the MS is explained by the fixed effects. The intra-subject variability in the intercept and the gradient is induced in the model by the random effects of TD, PD, $\Delta D$ factored by individual subjects.  The marginal and combined r-squared values are obtained using the `r2glmm' package. Both fixed and random effects combined accounted for approximately 96.34\% (95.91\% for RI) of variability in the match score. The 3.6 - 4.1\% (approx.) variability which is not explained by any covariates in the model can be directed to the miscellaneous factors like illumination and physiological condition which is beyond the scope of this study. Restricted or Residual Maximum Likelihood (REML) method is applied for model fitting. The model is created based on 10096 observations from RI and 10332 observations from LI grouped into 209 distinct individuals. The fixed and random effects are summarized in Table~\ref{tab:fixed_effect} and Table~\ref{table:RE} respectively. 
 
\begin{table}[!t]
\footnotesize
    \centering
    \caption{\textbf{Fixed Effects for Left and Right Iris}}
    \begin{tabular}{|c|c|c|c|}
    \hline
        \textbf{Variable} & \textbf{Parameter} &  \vtop{\hbox{\strut \textbf{Left Iris}}\hbox{\strut (\textbf{Est $ \pm $ SE)}}}  & \vtop{\hbox{\strut \textbf{Right Iris}}\hbox{\strut \textbf{(Est $ \pm $ SE)}}} \\
        \hline
        \hline
         Intercept & $\beta_{0}$ & 387.0$\pm$ 12.7 ***  & 401.5 $\pm$ 11.6 ***\\
         \hline
         TD & $\beta_{1}$ & -1.09 $\pm$ 0.33 ** & -1.39 $\pm$ 0.31***\\
         \hline
         EA & $\beta_{2}$ & 10.88 $\pm$ 3.4 ** & 13.3 $\pm$ 3.1*** \\
         \hline
         PD & $\beta_{3}$ & -0.74 $\pm$ 1.3 NS & -1.4 $\pm$ 1.3 NS\\
         \hline
        $\Delta D$ & $\beta_{4}$ & -394.9 $\pm$ 78.9*** & -269.5 $\pm$ 61.86***\\
        \hline
        $EA^2$ & $\beta_{5}$ & 1.4 $\pm$ 1.7 NS & -1.9 $\pm$ 1.5 NS\\
        \hline
        $PD^2$ & $\beta_{6}$ & -0.22 $\pm$ 0.20 NS & -168.2 $\pm$ 0.16 NS\\
        \hline
        $\Delta D^2 $ & $\beta_{7}$ & -1540 $\pm$ 334.9 *** & -1360.2 $\pm$ 355.7***\\
        \hline
        \multicolumn{4}{c}{Significance Code: 0 `***' 0.001 `**' 0.01 `*' 0.05 `.' 0.1 ` ' 1 ; *** indicates}\\
        \multicolumn{4}{c}{ p-value between 0 and 0.001 with significance level 0.001 and so on.}\\
        \multicolumn{4}{c}{Est.: Estimate, SE: Standard Error, NS: Not significant}\\

    \end{tabular}
   
    \label{tab:fixed_effect}
\end{table}

\begin{table}[!t]
\scriptsize
\centering

\caption{\textbf{Random Effects Left and Right  Iris}}
\begin{tabular}{|*{4}{c|}}
\hline
\multirow{2}{*}{\textbf{Groups}}  &  \multirow{2}{*}{\textbf{Parameter}} & \multicolumn{2}{c|}{\textbf{Standard Deviation}} \\
 \cline{3-4}
&  & Left Iris & Right Iris \\
\hline
\hline
 Intercept & $b_{0i}$  & 129.7 & 118.9 \\
\hline
 TD & $b_{1i}$  &  4.2 & 3.9 \\
\hline
PD & $b_{3i}$ & 15.7 & 15.3 \\
\hline
$\Delta D$ & $b_{4i}$ & 886.38 & 636.08 \\
\hline
$PD^2$ & $b_{6i}$ & 2.4  & 1.8 \\
\hline
Residual &   &  40.7 & 40.6 \\
\hline
\end{tabular}
\label{table:RE}
\end{table}

The analytic interpretation of the output of the model from G1-LI is noted below. G1-RI has the same interpretation, but is not included due to space limitations.

\begin{itemize}[leftmargin=*, noitemsep]
 \item Null hypothesis ($H_{0}$): There is no correlation between the individual predictors and the response variable (MS).
 
 \item Alternative hypothesis ($H_{A}$): There is correlation between the individual predictors and the response variable.  
 
 \item Estimate (Est.) provides a measure of change in MS from the intercept for each unit change in each variable. 
 
 \item Very high similarity is observed in both estimate and variability of TD coefficient ($\beta_{1}$) between both RI and LI. The estimated decrease in MS, $\beta_{0}$, due to TD estimate, $\beta_{1}$, is 1.09 $\pm$ 0.33  (approx) with each month increase in the time gap between mated pair of images, given that the null hypothesis is true. With p$<$0.01, the null hypothesis is rejected. This signifies there is a statistically significant decaying relationship between time gap and match score. However, the practical significance of the estimate is trivial considering the range of the scores (0 to 1557), the fixed effect intercept being 387.0 (401.5 for RI), and the threshold score being 36 at FAR 0.1\%. For example, a MS estimate of 387.0 would reduce to 347.76 in three years considering the effects predicted by the model. This score would still result in a match and does not impact biometric recognition over three years. 

 \item A positive baseline shift is noted for MS estimate, $\beta_{0}$, due to EA estimate, $\beta_{2}$. With an increase of EA by 1 year, the MS increases by 10.8 (13.3 in RI) $\pm$ 3 (approx) with p $<$ 0.01. This implies that EA has a significant effect on MS. So, with higher EA, a higher MS can be noted. It is expected that with increased age the interaction of the children with the data acquisition system improves. Older kids are able to hold still and have higher stability by keeping their eyes open and directed towards the sensor until the sensor auto-captures. The effect of EA on MS is linear as the quadratic factor is statistically insignificant. 
 
 \item Pupil dilation of the probe image (PD), $\beta_{3}$  affects the MS negatively; however the impact is statistically insignificant. 
 
 \item Difference in pupil dilation ($\Delta D$) between two mated pair of images, $\beta_{4}$, has negative correlation with MS. $\Delta D$ in our dataset varies in the range of 0.1 to 0.4. Thus the MS may decrease in the range 39.4 to 157.6 due to $\Delta D$. It is important to note that the effect on MS due to $\Delta D$ is higher by approximately 29 to 45 times than that of TD (LI). The effect of $\Delta D$ is non-linear. 
 
 \item  Ordering covariates based on adding random variability to the response in the model while considering intercept for each individual subject (random effect) as seen in Table~\ref{table:RE}: \\  $\Delta D$ $>$ Subject $>$ Dilation of the probe image $>$ Time Difference \\  \textit{Interestingly, we can infer from the model that dilation difference between two comparing images (in our case from 2 different TFs) adds in the most variability to the match score than any other significant factor. Aging effect contributes to the least variability in MS.}
\end{itemize}
 
\subsection{Iris Match Score Analysis: Age Based Groups}\label{Age_group}
\begin{figure}[!t]
\centering
    \includegraphics[width=3.2in, height=1.8in]{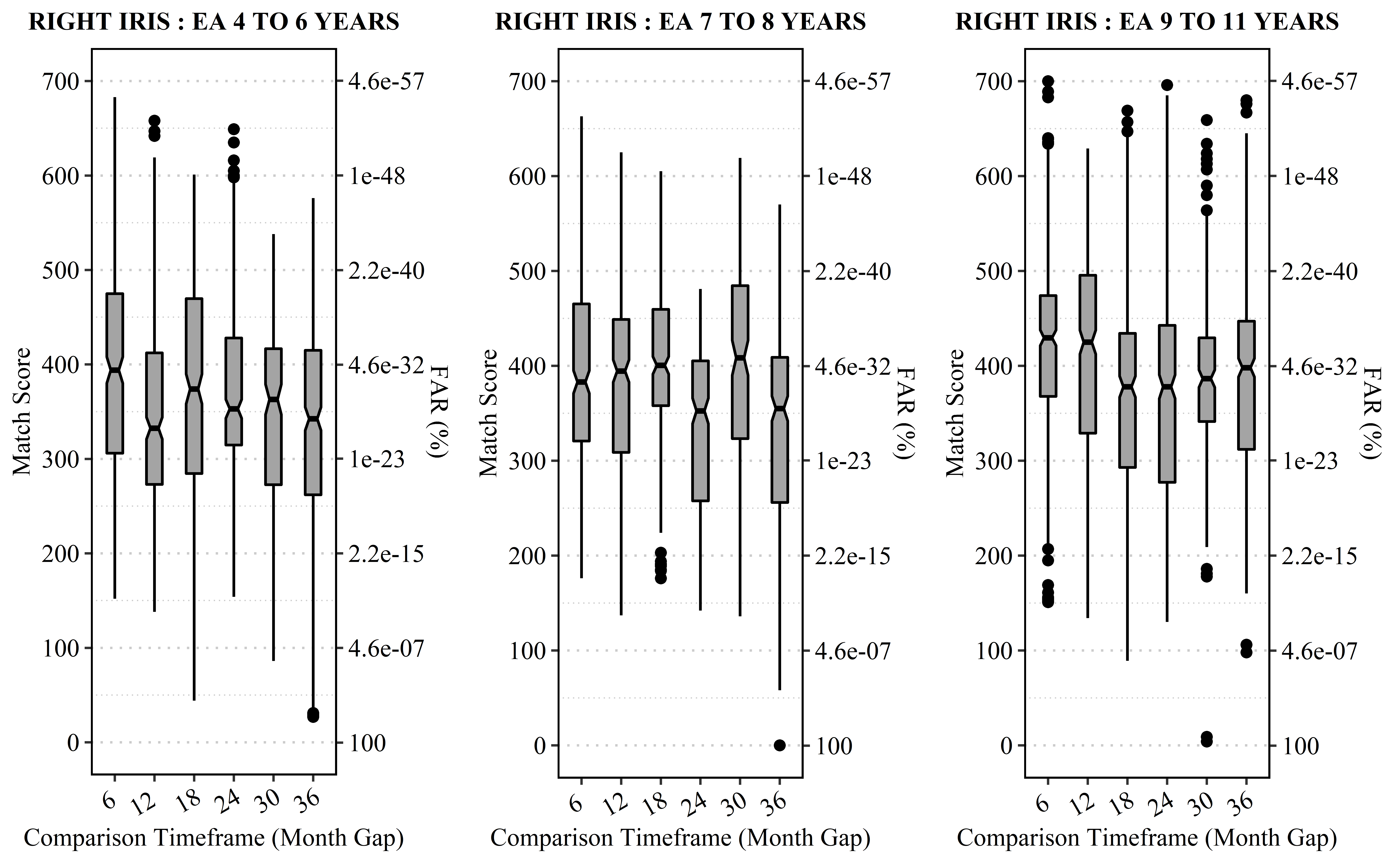}  
    \caption{\footnotesize Box-plots of all scores from subjects grouped based on age for G2 (all subjects have mated pairs from at least at 6 and 36 month TF) for different TFs for RI. G1 and G3 plots are similar to this and thus are not included to accommodate space and reduce redundancy. Relative False Accept Rate (FAR) w.r.t the MS is shown in the right y-axis.}.
    \label{fig:age_boxplot}
\end{figure}

The dataset has subjects aged between four to 11 years at enrollment. For this analysis, the entire dataset is grouped into three sections based on enrollment age - 4 to 6 years, 7 to 8 years and 9 to 11 years for analysis. The count of subjects for each age group are non-uniform; bin sizes are summarized in Table~\ref{table:Bin}. Each age group has been analyzed for each participation based group - G1, G2 and G3 following the LMER model represented in equation~\ref{eq:Full Model}. We do not have age information from 2 subjects in G1 and thus were removed. Thus, instead of 209, 207 subjects were analyzed in this section for G1. Boxplots of MS as a factor of TF for different age groups are shown in Figure~\ref{fig:age_boxplot}. The attributes of the boxplots for G1, G2 and G3 are similar; thus only plots from G2 are presented to accommodate space. Conclusion from the LMER model for each age group are noted below:
\begin{itemize}[leftmargin=*, noitemsep]
\item Estimated TD has a similar impact on MS on different age groups as our previous model. With the estimated decay of MS $\beta_{1}$ by $<$ 1.02 (p$<$ 0.01) for each month difference between the capture of the mated pair of images, the impact is practically insignificant. For the age group of 4 to 6 years, TD is not a statistically significant predictor of MS.

\item $\Delta D$ has similar correlation with MS as we noted in the previous model analysis. The impact is linear for the 4 to 6 years age group and is non-linear for the other age groups.

\item In contrast to our observation from group-based analysis, where all age groups were modeled as a whole, EA is not statistically significant when considering shorter age ranges compared to a larger age range of the full analysis.
\end{itemize}

We also did t-test analysis to test the difference in mean MS between different age groups, if any. The test concluded that the mean MS of the age group 9 to 11 is higher than 4 to 6 years (p$<$0.001) and 7 to 8 years (p$<$0.01). The difference in mean MS between age groups 4 to 6 years and 7 to 8 years is not statistically significant (p$>$0.05). However, we do not notice any impact on the tails of the distribution i.e the iris performance is not impacted as a result of age or aging as will be discussed in future sections. However to make a substantial conclusion on this, a more intensive study for each age is needed.

\begin{table}[!t]
\scriptsize
\centering
\caption{\textbf{Number of Subjects for Each Age Group}}
\label{table:Bin}
\begin{tabular}{|c|c|c|c|c|}
\hline
\textbf{Groups} & \textbf{\makecell{4 to 6 years\\ (LI/RI)}}  & \textbf{\makecell{7 to 8 years\\ (LI/RI)}} & \textbf{\makecell{9 to 11 years\\ (LI/RI)}} & \textbf {Total} \\
\hline
\hline
\textbf{G1} & 90/90 & 54/54 & 63/63 & 207/207\\
\hline
\textbf{G2} & 25/25 & 37/34 & 43/42 & 105/101 \\
\hline
\textbf{G3} & 13/15 & 22/21 & 28/26 & 63/62\\
\hline

\end{tabular}
\end{table}

\subsection{Dilation and its Relationship with Match Score}\label{Dilation}
Dilation is an important factor to be considered in iris recognition. Studies have confirmed change in ocular biometric parameters with change in dilation \cite{pupildilation2018}. These changes might affect the performance of algorithms used to generate a match score. Variable dilation is considered as an explanatory variable in the match score variability with time in recent research concerning evaluation of aging effect \cite{grother2013irex}. Physiologically, studies have indicated that pupil size varies with age \cite{adler1965physiology}. Pupil size is small for the first year after birth and gradually attains its maximum size in childhood and adolescence and again gradually becomes smaller with time. However, the pattern may not be reflected in this study due to the limited TF of 3 yrs and having no infants in the dataset. Dilation is also a response variable to environmental factors like illumination, weather, subject's medical history and fight and flight response of the Central Nervous System in response to a stressful environment. In this section, we study dilation as an independent factor impacting performance of iris recognition. Dilation is derived as defined in Equation~\ref{eq:D}.\par
\begin{figure}[!t]
\centering
\subfigure{(a)\label{subfig:coll_vs_dilation}}{\includegraphics[width=4.4cm, height=5cm]{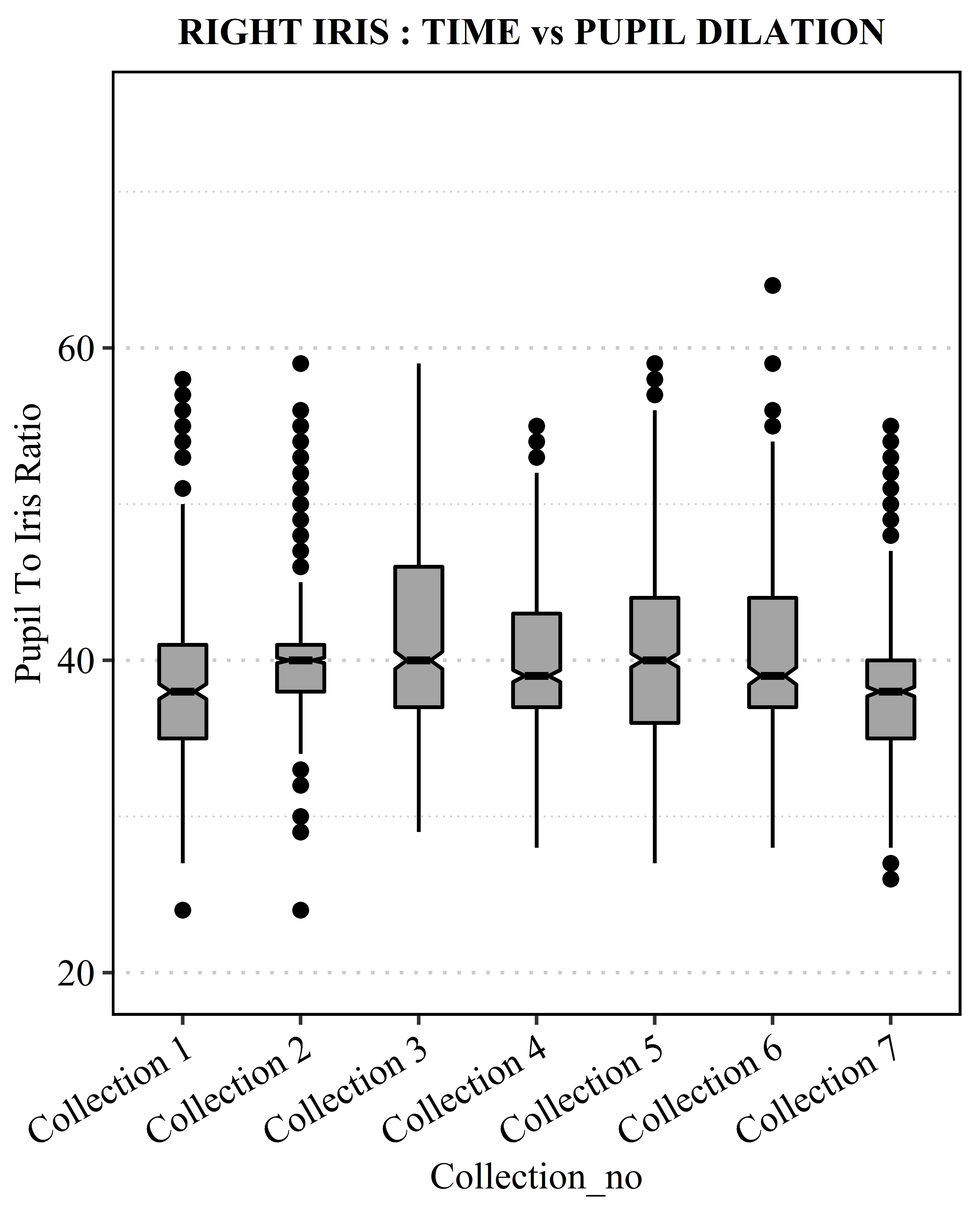}}
\hfill
\subfigure{(b)\label{subfig:time_vs_dilation_constancy}} {\includegraphics[width=3.5cm, height=5cm]{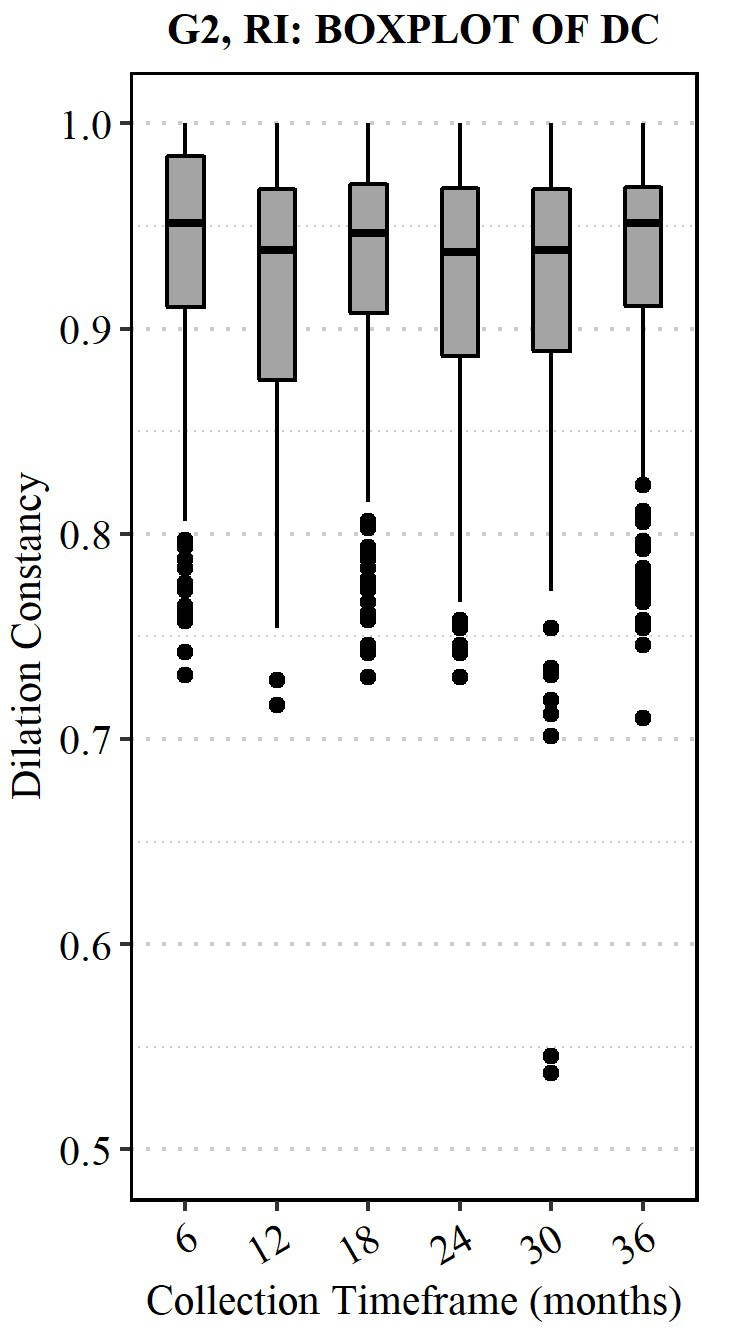}}
\caption{\footnotesize (a)  Boxplot of pupil dilation from all images captured in each session for RI ; (b) Summarized iris dilation constancy compared to the enrollment image at different TFs (6, 12, 18, 24, 30, 36 months from enrollment) for right iris }
\end{figure}

 
Figure~\ref{subfig:coll_vs_dilation} shows dilation for each collection which includes all participating children at a fixed time point (e.g. Collection 2 in November 2017). All statistical tests were compared to Collection 1. The t-test of the mean dilation for the six collections compared to Collection 1 shows that the mean dilation is relatively constant (p$>$0.05) except for Collection 3 (p$<$0.01). Chi-square analysis of dilation concludes that there is statistically significant (p$<$0.05) difference in the dilation variance of Collection 3, 5, 6 and 7 when compared to Collection 1. Measures are taken during collection to minimize impact of illumination and weather as detailed in Section~\ref{Methodology} and \ref{iris_coll}.  Although  measures are taken to control data collection environment, the dataset is not immune to impact from miscellaneous variability factors.\par
\begin{figure}[!t]
\centering
    \includegraphics[width=3.5in]{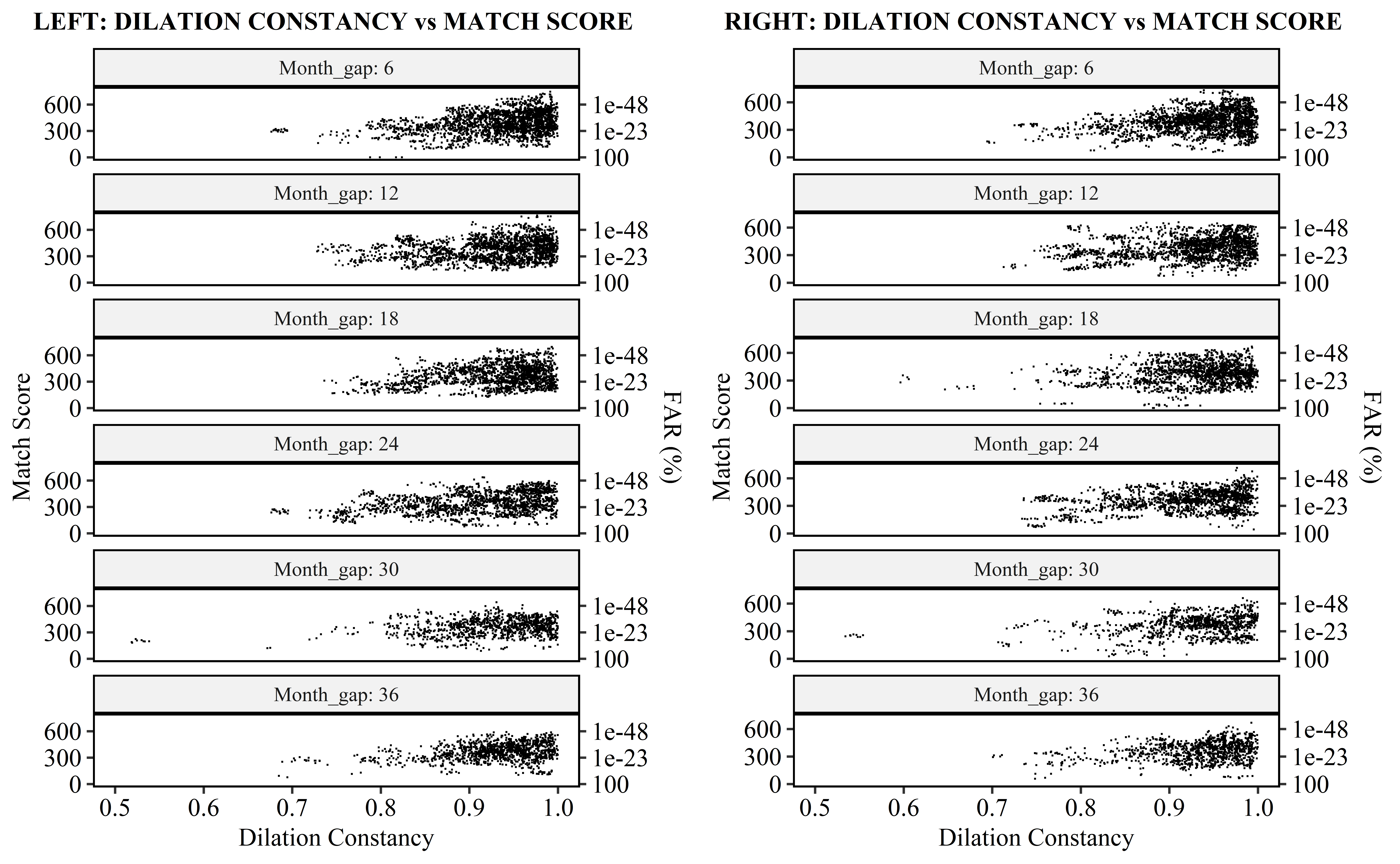}
    \caption{\footnotesize Graphical presentation of match scores (y-axis) and dilation constancy (x-axis) for different time-periods (top to bottom: 6, 12, 18, 24, 30, 36) for left and right iris (left and right)}
    \label{fig:dilation_MS}
 \end{figure}

 Dilation constancy (DC) is the measure of similarity in pupil dilation between enrollment and probe image as detailed in Section \ref{sec:DC}. Figure~\ref{subfig:time_vs_dilation_constancy} summarizes the DC between enrollment image and respective probe images collected at different TFs. Only the graph from G2-RI is included in the paper to accommodate space. All three groups based on participation for both LI and RI show similar attributes. In the majority of the TFs across different groups, the upper 75\% of the mated pairs have DC above 0.9. The median DC is greater than 0.9 across all sessions and all TFs. In our dataset, the DC ranges from 0.55 to 1 and we observe more outliers in the lower quartile of the plots. We implemented a LMER model as described in Equation~\ref{eq:DC Model} to predict DC employing TF and EA as predictors. From the model we conclude that the impact of TF and EA on DC is statistically insignificant (p$>$0.05). \textit{Increased time difference between probe and gallery does not influence the DC.}   \par

\begin{equation}\label{eq:DC Model}
\small
    \mathit{\mathbf{DC \sim  \boldsymbol{\beta}_{0} + \boldsymbol{\beta}_{1}TD  +\boldsymbol{\beta}_{2} EA+ \beta_{5} EA^2 + b_{0i}+ b_{1i}TD  }} 
\end{equation}
where,
\begin{itemize}
 
\item $\beta_{k}$ and $b_{ki}$ are the fixed regression coefficient for corresponding parameter, \textit{k} and the random regression coefficient for corresponding parameter, \textit{k}, for subject,\textit{i}, respectively.
  
\item $\beta_{0}+ b_{0i}$ is the sum of fixed and subject specific random intercept corresponding to the initial state.
\end{itemize}
Figure~\ref{fig:dilation_MS} illustrates the relationship between MS and DC. A triangular pattern is noted. Lower dilation constancy translates to lower match scores; however, low MS can be seen even with high dilation constancy likely due to other factors such as usable iris area, noise, or other quality factors. DC, as has been previously reported, impacts the MS. However, in this study, the Pearson's correlation between DC and MS of all matches performed is weak with correlation coefficient of 0.28. DC in this dataset, with substantial outlier variability has not impacted the tails of the distribution i.e. \textit{DC did not impact practical application of iris recognition in children to a point of false non-match}. Saying that, iris recognition performance is sensitive to dilation and thus the data has been collected in a semi-controlled environment to minimize impact on dilation. \par

 \subsection{Iris Recognition Performance: False Non-Match Rate}\label{FNMR}
 False Non-Match Rate (FNMR) is an evaluation metric used in prior aging studies as discussed in Section~\ref{SOA} and defined in ISO/IEC standard 2382-37\cite{ISO_2382_37}; it is the fraction of mated (from the same individual) images that fail to match; it has operational impact. We calculated FNMR with a MS threshold value of 36, equivalent to False Match Rate (FMR) of 0.1\% using the VeriEye FMR calibration\cite{VerieyeSDK}. Images were rejected when the MS were below the set threshold. A chronic increase in FNMR with time may be correlated with degrading biometric performance with increased time duration between enrollment and probe.\par
 
\begin{figure}[!t]
\centering
    \includegraphics[width=3.5in]{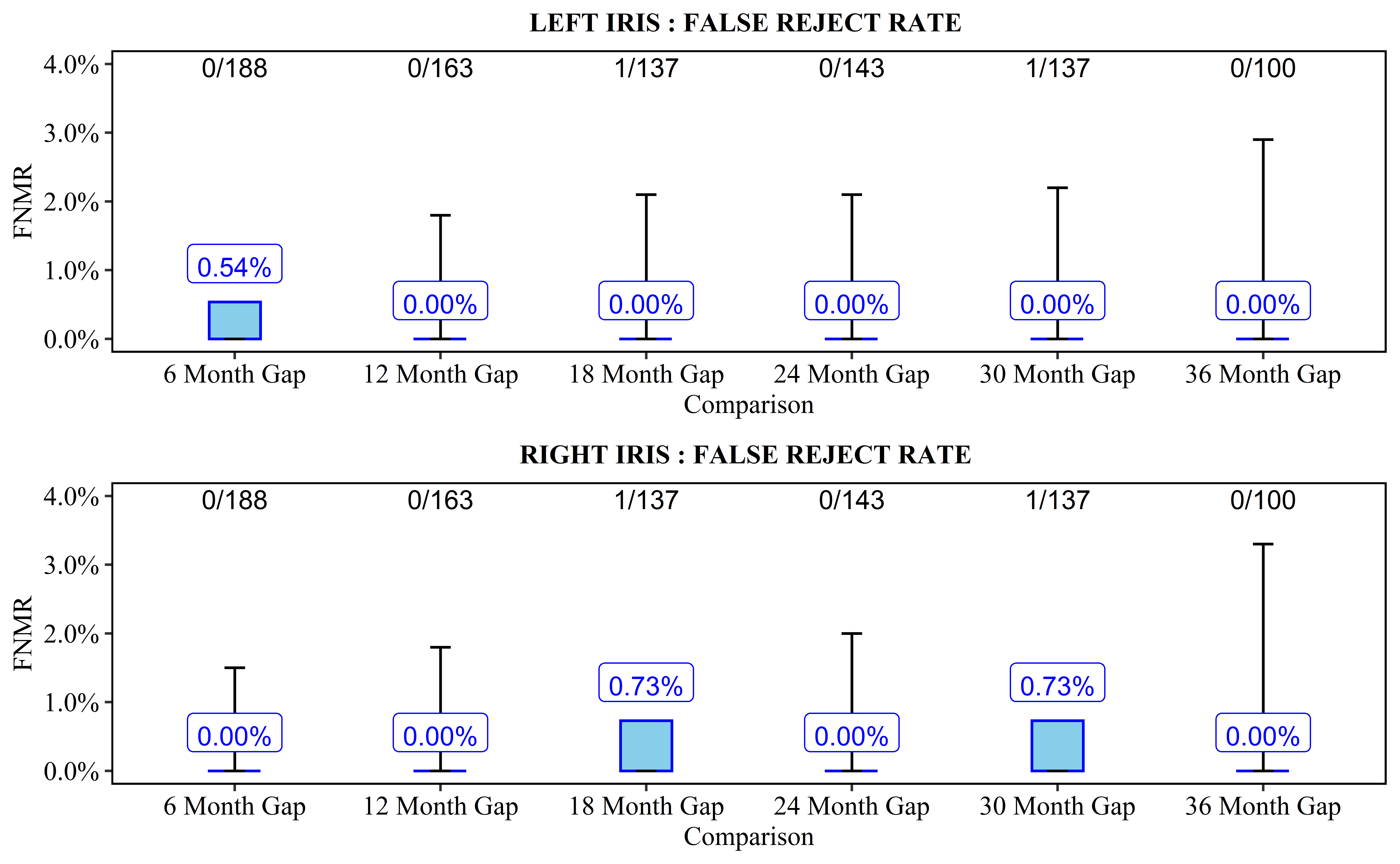}
    \caption{\footnotesize FNMR at different TFs (6, 12, 18, 24, 30 and 36 months from enrollment) for left and right iris. Values marked in blue indicate FNMR based on our observation at 0.1\% FMR. The black lines show the error-bars calculated based on 'Rule of 3'. X/Y above each bar indicates the number of subjects rejected/total number of subjects for this TF}
    \label{fig:FNMR}
 \end{figure}

 Variation in FNMR is shown in Figure~\ref{fig:FNMR}. Rejection is observed only in G1. No rejections were seen from G2 and G3. Single cases of rejections are observed at 6, 18 and 30 month TF. The three occurrences of FNM cases are summarized in Table~\ref{table:FR}. At 6 month TF, RI of one of the 186 participating subjects is rejected; thus, the FNMR at 6 months is 0.54\%. At each of 18 and 30 month TFs, the LI of one of the 137 participating subjects were rejected; The FNMR at both TF is 0.73\%. The slight increase in the FNMR from 0.54\% to 0.76\% is not an increasing trend in the FNMR but a reflection of decreased number of subjects. \par  
 
\begin{figure}[!t]
\centering
    \includegraphics[width=3.5in]{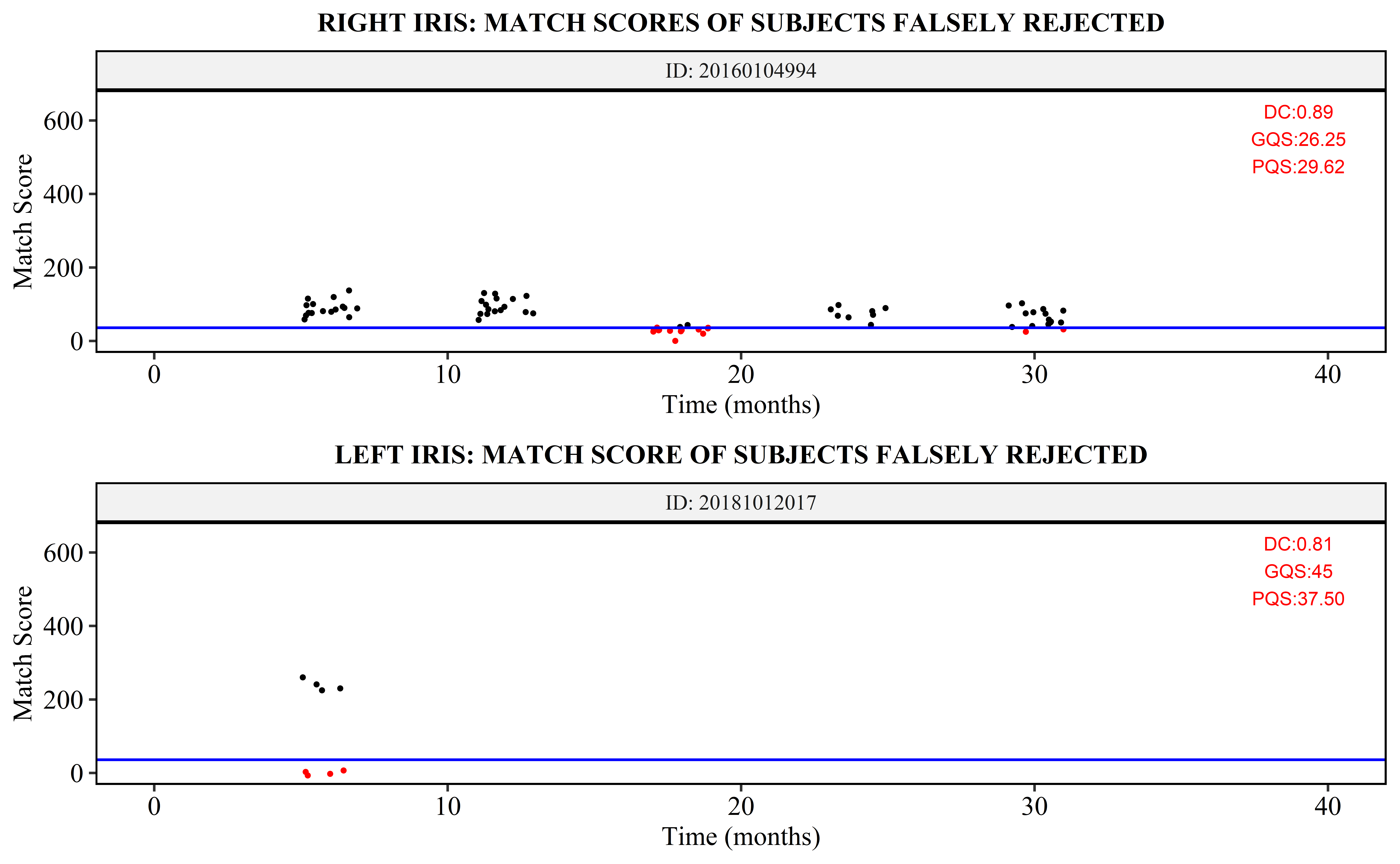}
    \caption{\footnotesize Match scores for all cases of rejections in the study. Each graph shows the match scores (y-axis) of individual subjects at different time-frames (x-axis). The rejected comparisons MS are marked by red dots. The blue line indicates the set threshold (MS 36 at 0.1\% FAR) for the study. `DC', 'GQS' and 'PQS' denotes the average dilation constancy, average quality score of gallery and average quality score of probe of all rejected images of that subject respectively}
    \label{fig:FR_MS}
 \end{figure}

 \textit{ Out of 20428 mated pairs of images from 418 different irides in the study 20408 mated pairs were above the threshold and results in a match.} 20 mated pairs from two different irides from two subjects failed the set criteria and were falsely rejected. The failures are summarized in Table~\ref{table:FR} and detailed analysis and understanding of the root causes are elaborated in Section~\ref{Detailed_FNM}. Subject 20181012017 had rejection in the LI in one session and subject 20160104994 in the RI in two different sessions. No subjects were rejected in both irides. When a subject is rejected, not all images from that session are rejected (refer to Table~\ref{table:FR} - Number of Rejections). The average DC of the rejected images is substantially high (0.8 to 0.9). Thus the FNM are not a result of a large dilation difference between the mated pair of images in the dataset. Since multiple images from a single triggered capture were collected on each visit (except seventh session), the images have high correlation. Due to the intrinsic setup of the sensor, the sensor provides multiple images for each triggered capture, all captured within a few seconds. However, with change in protocol in the 7th session, multiple images were collected in two sets, decreasing the high correlation between images collected in different sets. Two of the three occurrences of rejections were collected in the 7th session (refer Table~\ref{table:FR}). Figure~\ref{fig:FR_MS} depicts the individual exploratory view of the MS from the two subjects for all the TFs they participated. Subject 20181012017 has recorded comparisons at 6 month TF and showed rejections in four of the eight mated pairs. Since all of the images were not rejected it is highly unlikely that the rejection is due to irreversible change in the iris texture. Subject 20160104994 had matching data from five TFs (6, 12, 18, 24 and 30 month TF) in which rejection was observed in two sessions - 18 and 30 months. At 18 month TF, 14 of the 16 mated pairs were rejected with the other two having scores just above the threshold. At 30 month TF only two of the 16 mated pairs were rejected. However the MS of the matched pairs of comparison exceed the set threshold by a large margin (refer Figure~\ref{fig:FR_MS}). In between 18 and 36 month TF, the MS "returns" to above threshold at 24 month TF. These sporadic false non-matches and a trending low MS all through the five sessions indicate that the false non-matches are a result of other factors than aging. The underlying causes of rejections are explored in more detail in the next section.

 \begin{table*}[!t]
\scriptsize
\centering
\caption{\textbf{Summarizing the False Non-Matches in the Study}}
\label{table:FR}
\begin{tabular}{|c|c|c|c|c|c|}
\hline
\textbf{Coded ID} & \textbf{Collection Number for Rejection}  & \textbf{Time-frame of Rejection} & \textbf{Number of Rejections} & \textbf{Nature of Failure} & \textbf{LI/RI/Both} \\
\hline
\hline
20181012017  & Collection 7 &  6 month & 4 of 8 mated pairs   & \makecell{Angle of Presentation \\ w.r.t. camera } & LI\\
\hline
20160104994 & Collection 5 and 7 & 18 months; 30 months & \makecell{18 months:  14 of 16 mated pairs \\ 30 month: 2 of 16 mated pairs} & Poor Enrollment & RI\\

\hline

\end{tabular}
\end{table*}

 Since the occurrences of false matches at different TFs are episodic, the results of the FNMR as marked in 'blue' in Figure~\ref{fig:FNMR} are statistically analyzed by the `Rule of 3' and the projected FNMR are marked by the error bars. This rule has been adapted in practice in biometrics and is described in ISO/IEC 19795-1\cite{iso2006iec}. The rate of occurrence of a particular event in a population which does not occur in an experiment is calculated by the `Rule of 3'. In our study, at particular TFs (12, 24 and 36) there is no occurrence of FNM. The rule states that 3/N is the upper bound of such an occurrence at 95\% confidence interval, where N is the number of people in the experiment. We have 209 subjects in the study. Thus based on this rule the range of FNMR in this study is 0\% to 1.4\%. However, since our experiment is longitudinal, having different numbers of subjects at different TFs and between LI and RI, the fragmented projected maximum FNMR is shown in Figure~\ref{fig:FNMR}. The maximum upper bound FNMR is 3.3\% for RI at 36 months. Instances where the projected error is not shown are cases where we have recorded false non-matches and do not qualify the criteria for the `Rule of 3'. The increase of the upper limit of FNMR is a reflection of the decreasing number of subjects, i.e., the upper bound is higher when there are fewer subjects.
 
 \subsection{Detailed Examination of FNM Images}\label{Detailed_FNM}
 We did a root cause analysis for false non-matches of the two subjects (refer Table~\ref{table:FR}) individually.

\subsubsection{Case 1: Coded ID - 20181012017}
\begin{figure}[!t]
\centering
    \includegraphics[width=3.5in]{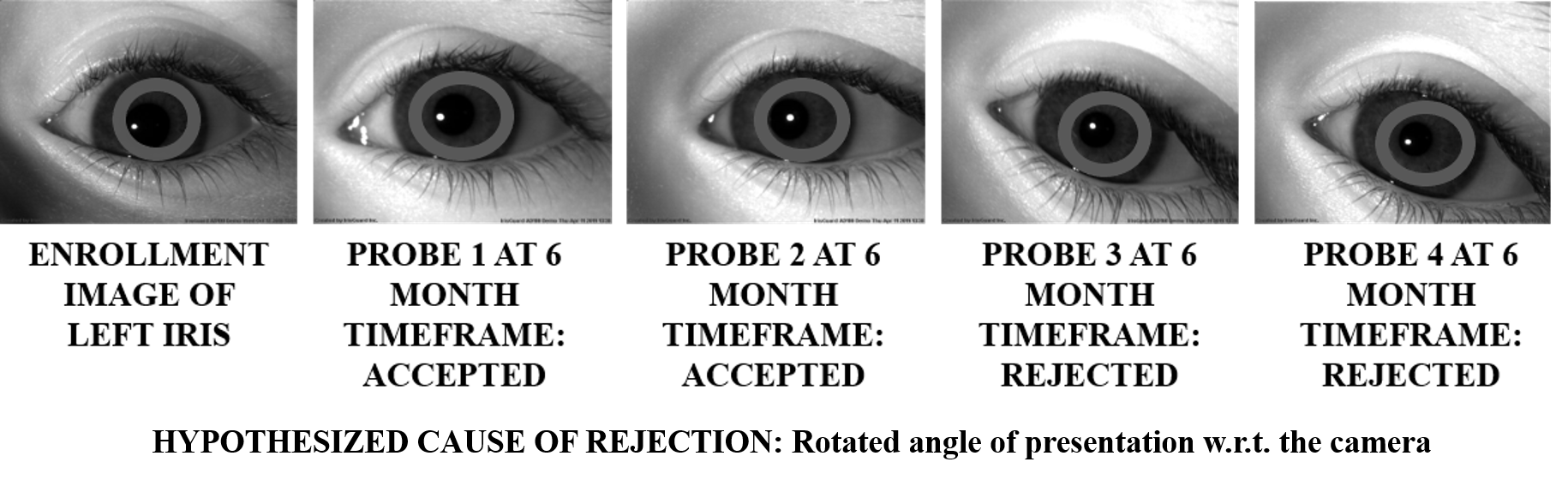}
    \caption{\footnotesize False Non Match: Case 1: Iris images of subject 20181012017 at enrollment (left) and the probe images matched and falsely non-matched at 6 month TF. Iris patterns are partly obscured to maintain privacy.}
    \label{fig:Rejection_017}
 \end{figure}

The subject was enrolled in the 6th session, having two collections in the dataset i.e one recorded TF at 6 month.  Rejection was observed only in the LI. Four out of eight comparisons (2 of 4 images from the 7th session) are rejected. The 4 comparisons from the truly accepted mated pair had an average MS of 238. The rejected 4 comparisons had an average MS of 0. The MS at different TFs are plotted in Figure~\ref{fig:FR_MS}. The average DC of the four rejected pair of mated images is 0.81. The average overall image quality of the rejected images of the gallery images (GQS) is 45 and the probe images (PQS) is 37.5 which are below the mean overall quality score distribution. The images are displayed in Figure~\ref{fig:Rejection_017}. 

Visual inspection revealed that in the two rejected iris images (4th and 5th images in Figure~\ref{fig:Rejection_017}), the eye is captured at a rotated angle of 15 degree w.r.t the camera. VeriEye being a black box, we are not aware of the compensation adapted by the algorithm to accommodate  angle of rotation considered in their algorithm. However, manually rotating the images by -1 degree, the VeriEye software is able to match the images. This is a case where the issue is the presentation of the iris to the camera which is challenging in children and has been observed in previous studies \cite{basak2017multimodal} \cite{nelufule2019image}. We conclude that the root cause of rejection has no relation to age/aging of the iris biometric. However, errors such as these lead to false non-matches. \textit{We observe and suggest that when working with children there is a need for a tool to detect the canthi and rotate the image to bring the canthi onto a horizontal line for cases where the image has significant rotation.}  

\subsubsection{Case 2: Coded ID - 20160104994}
The subject participated in six of the seven collections. We have matching information from five TFs (6, 12, 18, 24 and 30 months). The MS at different TFs are plotted in Figure~\ref{fig:FR_MS}. The subject was falsely rejected at 18 month and 30 month TF only in the RI. At 18 month TF, 12 of the 16 comparisons performed are rejected. The average MS of the four matched image-pairs is 38.2, which is just above the set threshold. At 30 month TF, 2 of 16 comparisons performed are rejected. The average MS of the 14 matched pairs is 66.2. As noted, the subject has a very low MS for all true matches across different TFs. The average DC of the rejected mated pairs of images is 0.89.

\begin{figure}[!t]
\centering
    \includegraphics[width=3.5in]{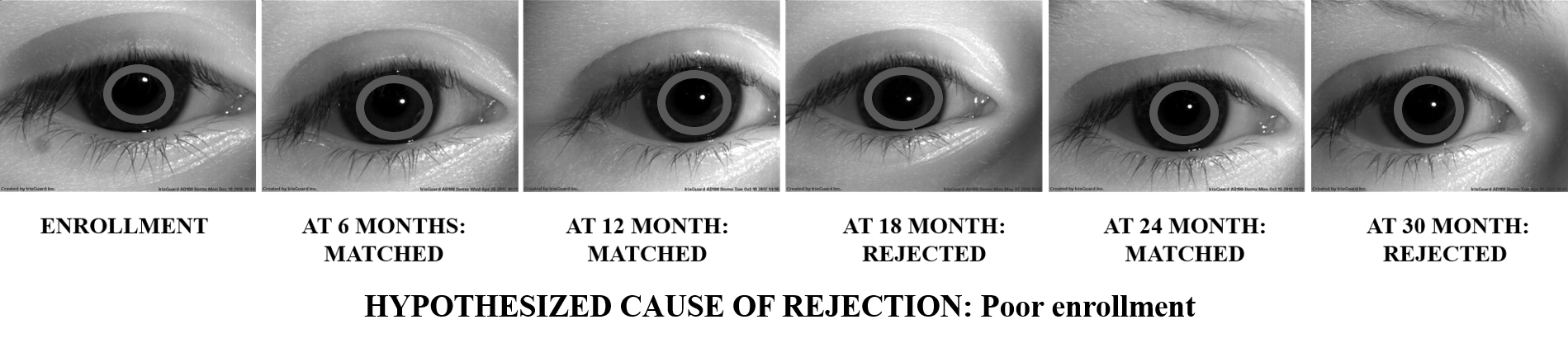}
    \caption{\footnotesize False Non Match: Case 2: Example of images collected at different TF for subject 20160104994. The subject matched with the enrollment image at 6 month and 24 month TF but falsely non-matched at 12 and 30 month TF}
    \label{fig:Rejection_994}
 \end{figure}
 
 Visual inspection and analyzing the quality factors of the images from the subject renders the following observations:
 
 \begin{itemize}[leftmargin=*, noitemsep]
    \item Visually, the enrollment image has "poor quality" in terms of the exposed iris area, occluded iris with eyelash and low contrast between eyelash, pupil and iris.
    
    \item Visually, the contrast between the iris and the pupil is substantially low even with NIR illumination for this subject across all TFs, giving a salient notion of a "dark iris". 
     \item The average overall image quality of the rejected images of the gallery images(GQS) is 26.25 and the rejected probe images (PQS) is 29.62 on a range of 0 to 100. The average usable iris area of the 4 enrollment images is 59.75\% and the average pupil to iris contrast is 52.7. All 3 quality measures - overall image quality, usable iris area and pupil to iris contrast of the rejected images falls in the lowest quartile of the distribution.  
\end{itemize}
  
Since the MS across all TFs for this subject are substantially low, we are inclined to believe that this is a case of poor enrollment. To further explore this, we set aside the images from the original enrollment session and considered the subsequent collection (the subject's second session) as the enrollment and match it against all subsequent collections. All mated pairs match. Thus we conclude that the rejections observed in this case is due to poor image quality of the enrolled image and is not affected by time-frame.

\subsection{FMR and FNMR}
\begin{figure}[!t]
\centering
\includegraphics[scale=0.3]{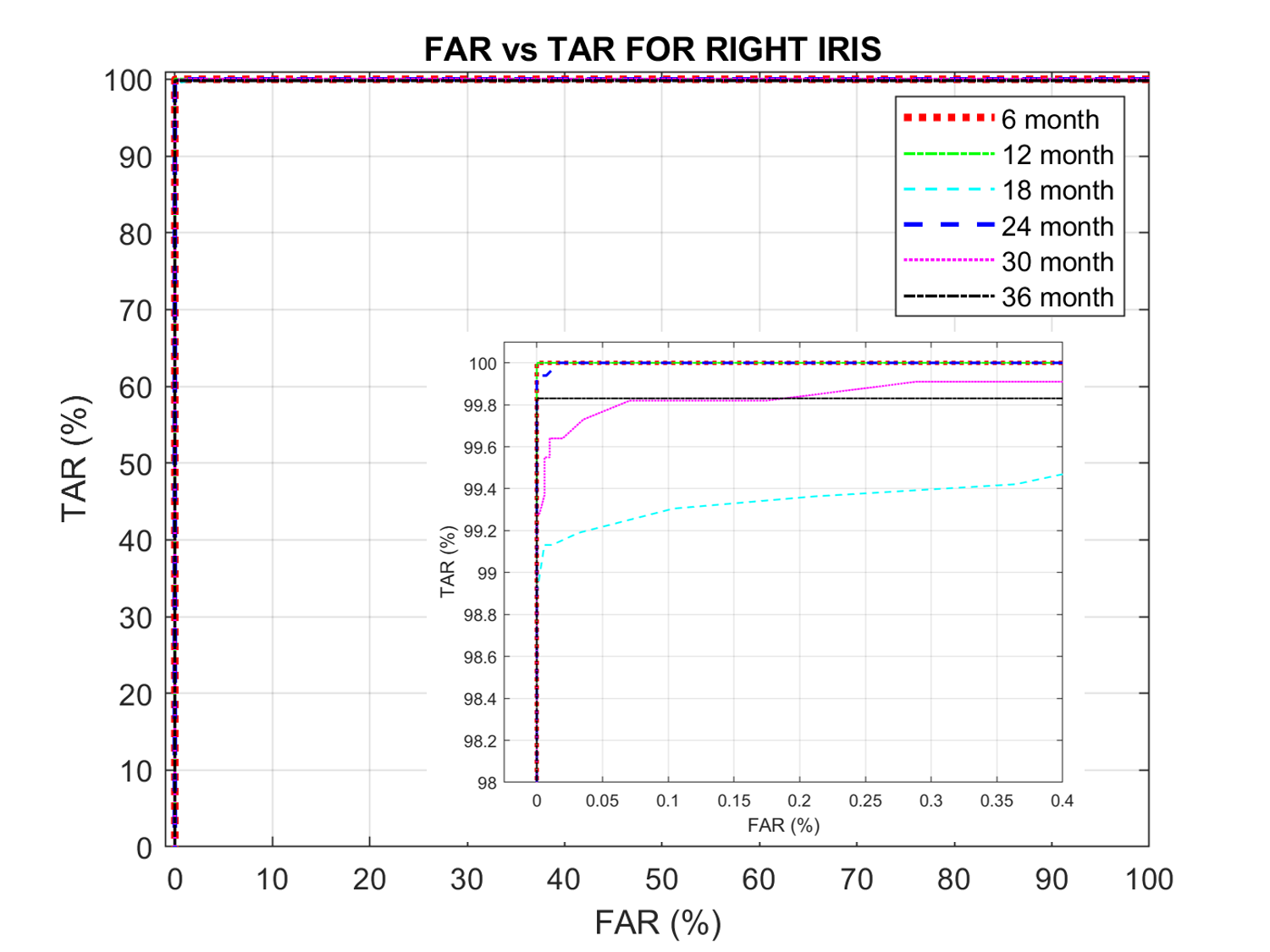}
    \caption{\footnotesize ROC curves for different TFs; Inset image is a magnified representation}
    \label{fig:FAR_TAR}
 \end{figure}

This section provides a more detailed analysis of ROC, False Match Rate (FMR) and False Non-Match Rate (FNMR) for our dataset. All reports are for RI. LI shows a similar pattern and is thus not included here. The ROC curve is shown in Figure~\ref{fig:FAR_TAR}. Table~\ref{table:FMR_FNMR_TH} compiles the FMR and FNMR over the 36 month TF at different  threshold  values. Verieye claimed a FAR of 0.1\% at threshold value 36. At that threshold, for our dataset the FMR varies between 0.03\% to 0.08\% at different TFs. 
   
\begin{table}[!t]
\scriptsize
\centering
\caption{\textbf{FMR and FNMR in \% at Fixed Thresholds}}
\begin{tabular}{ |p{1cm}|c|c|c|c|c|c| } \hline
 \textbf{Time \newline Frame} & \multicolumn{2}{c}{\textbf{Threshold 36}} & \multicolumn{2}{|c}{\textbf{Threshold 50}} & \multicolumn{2}{|c|}{\textbf{Threshold 100}} \\
 \cline{2-7}

  & \textbf{FMR} & \textbf{FNMR} & \textbf{FMR} & \textbf{FNMR} & \textbf{FMR} & \textbf{FNMR} \\
 \hline
 \hline
6 mos.  & 0.05 &	0 &	0 &	0 &	0 &	0.68 \\ 
 \hline
 12 mos. & 0.07 &	0 &	0 &	0 &	0 &	0.47 \\
 \hline
 18 mos.  & 0.03 &	0.81 &	0 &	1.16 &	0 &	1.4 \\
 \hline
 24 mos.  & 0.06 &	0 &	0.00 &	0.06 &	0 &	1.4 \\
\hline
30 mos.  & 0.07 &	0.18 &	0.01 &	0.63 &	0 &	1.8 \\
\hline
36 mos.  & 0.08 &	0.17 &	0.00 &	0.17 &	0 &	1.4 \\
\hline
\end{tabular}
\label{table:FMR_FNMR_TH}
\end{table}

\section{Discussion, Limitations and Future Work}\label{Limit}
Impact on iris biometric performance due to increasing time difference between gallery and probe of iris image in children is the focus of this work. Addressing this, iris images from 209 subjects in the age group of 3 to 14 years were collected in 7 sessions, spaced approximately six months over 3 years from the same subjects and were analyzed. In addition to the effect of time window between collections on match scores (MS), variability factors like dilation, dilation difference between enrollment and probe image, and enrollment age have been taken into consideration for analysis. The performance is assessed with false non-match rate (FNMR) and Linear Mixed Effects Modelling(LMER).\par 

 Statistical conclusions are drawn from the LMER (refer Section~\ref{LMER}) which modelled the match score (MS) variability as a function of increasing time difference between gallery and probe of iris image (TF) in presence of inter-subject and intra subject variability factors of enrollment age (EA), dilation (PD), and dilation difference between enrollment and probe ($\Delta D$). The model considers that the MS varies as a linear function of the predictor variables. However, non-linear impact of the predictors are taken into account by adding second order terms to the variability factors. With respect to the design of the model, a next step might be to consider other models to include additional order terms to fit this data. As Pinheiro and Bates\cite{pinheiro2006mixed} mentions \textit{"By increasing the order of a polynomial model, one can get increasingly accurate approximations to the true, usually nonlinear, regression function, within the observed range of the data''}. Additionally, non-linear models which also consider underlying mechanisms like asymptotes and monotonicity producing the data, while modelling response variables, could also be considered as future work. The same model is implemented for three groups based on participation - G1, G2 and G3. All groups render similar conclusions. The results can be summarized as: (i) a slight decay in MS (approx: 16.8/year and avg MS is 399) with increased TF in a three year period for the enrollment age of 4 to 14 years; this small decay has no practical impact on performance of iris recognition; (ii) strong negative correlation between $\Delta D$ and MS; (iii) no significant effect of PD on MS; (iv) linear significant effect of EA on MS ($ p < 0.01 $). Even with EA rounded off to the nearest year (age calculated from birth year), we see that EA is significant and thus believe important to consider it as a predictor in the model.\par
In addition to aging i.e increased time between enrollment and probe, studying the impact of age on MS is important to understand any underlying effect. We did a preliminary statistical assessment of the difference in mean MS between different age groups - 4 to 6 years, 7 to 8 years and 9 to 11 years. t-test statistics concluded that the mean MS of age group of 9 to 11 years is significantly (p$<$0.01) higher than other age groups. Additionally, we employed the same LMER model for 3 age based groups. The conclusions in all age based groups are mostly similar to the conclusions from LMER implementation on the participation based groups with few exceptions: (i) For 4 to 6 years age group, the time difference has statistically insignificant impact on MS and $\Delta D$ has linear impact on MS; (ii) Enrollment age is not a statistically significant predictor of MS for smaller fragmented age groups. Addressing age based issues is challenging due to limited data for each age group. The small, non-uniform count of subjects in each age for the age spread of 4-11 years from approximately 209 subjects poses a challenge to make conclusions. An age focused data set needs to be created to meet this challenge.\par

Dilation and dilation constancy (DC) between enrollment and probe is evaluated. DC varied between 0.55 and 1 in our dataset. Low DC necessarily translates to low MS; for high DC, MS varies from low to high (refer Figure \ref{fig:dilation_MS}), where low MS are likely due to other factors. The DC of the rejected images are substantially high (all above 0.8). Dilation difference between two mated pairs of images has a major contribution to errors in recognition performance. However, no case of rejection is recorded due to low DC in our database. In our study we note that the MS is weakly correlated to DC with Pearson's coefficient of 0.28. Dilation and DC impact the iris recognition on the MS independently.   Having very high/very low dilation in both enrollment and probe leads to a high DC. Very high dilation in both gallery and probe, though leads to a high DC, leaves a very low available iris area for analysis. A very low dilation may lead to segmentation error. We observe that both dilation and DC contribute to the MS with substantial effect. TD and EA have no statistical impact on DC. Dilation is a crucial factor and has multiple causes including age, illumination, medical causes and environmental factors. Of the studies reviewed in literature in Section \ref{SOA}, the closest to the age range of our study, were by Adler in \cite{adler1965physiology} which mentioned dilation reaching its maximum in the childhood and adolescence. Thus, a constructive physical relation between age and dilation remains unavailable. A comprehensive understanding of the relationship between age and aging on dilation in children based on our dataset was performed\cite{Das_dilaiton}. We were unable to acquire exact same dilation at each session from the same subject. However, even with this knowledge we are unable to have controlled constant illumination due to our dependency on the school setup who permits us to use their facility for collections. Measures including closing blinds and giving subjects time to adapt to the room, were taken to minimize the effect and variation of illumination and environmental factors. The Iris Guard camera \cite{IrisGuard} used for data collection, has a blinking white LED light with the purpose to provide external stimuli to accommodate pupil to an uniform dilation for each subject across sessions. Technology has been adapted by some software to accommodate dilation to minimize the dilation difference between collections. Statistical tests showed significant variability in dilation in collection sessions (Collection 3, 5, 6, 7) when compared to Collection 1. However, the mean dilation in different sessions remained relatively constant (exception: Collection 3 (p$<$0.01)) when compared to Collection 1. Illumination heavily impacts dilation. Thus it is essential to account for change in dilation and area of iris available for analysis while calculating the MS. \par

False Non-Match Rate (FNMR) in this study ranges between 0\% to 0.73\%. Two subjects were falsely rejected in the three sessions (6, 18 and 30 month TF).  The FNMR at 6 month is 0.54\% and 0.73\% at 18 and 30 months. The increase is not an indication of increased FNMR with time but a reflection of variable subject participation in different sessions. The FNMR being episodic with most sessions having 0\% FNMR, the upper bound (UB) of the confidence interval for FNMR is statistically calculated with the `Rule of 3'. The UB for FNMR in 36 month is projected at 3.3\%. FNMR in this study does not indicate aging effect on the iris recognition performance in children in the age group of 4 to 11 years at enrollment for a period of three years. Each individual case of rejection is studied in detail to understand the underlying cause as detailed in Section \ref{Detailed_FNM}. All cases of false non-matches were accounted for by factors encompassing poor enrollment and angle of rotation of the captured image w.r.t. to the camera. This observation goes to the heart of template aging questions. \textit{Even though changes in FNMR are seen, is this aging?} Indeed, false non-matches do not necessarily mean an irreversible change in the iris itself. In this study we verified that all rejected subjects returned to above threshold match scores, and identified the root causes for the rejections.\par

This work opens some major areas for future work based on the challenges and limitations faced. Further quality assessment is needed to understand the errors. Additional quality assessment of the iris images during capture would be a step towards accurate assessment of the data. Increased FNMR are primarily caused by poor quality images. Factors  may include partial occlusion of the area of interest, motion blur, unfocused image, high dilation or other quality issues. Standard Image Quality for iris is set in the report ISO/IEC 19794-6 by ISO (International Organization for Standardization) and IEC (International Electrotechnical Commission) \cite{iris_standard_report}. We plan to study quality in more detail in future study.\par

The biggest challenge in research is the limited availability of data, especially in the fields involving children. The dataset used in the study is an in-house dataset, which is part of an ongoing study. To the best of our knowledge this is the only longitudinal database in the focus age group of 4 to 14 years directed to study biometrics in children. Presently the study has enrolled 239 subjects. We believe analyzing longitudinal data from 209 subjects has statistical significance as we have reported in this paper. High correlation between images from the same time-frame is noted due to the collection protocol. All images were collected within seconds of each other based on the internal setup of the sensor for multiple image collection. The issue was identified and the protocol was modified from the 7th session. Multiple images are collected in 2 sets with approximately one minute in between. LI and RI images are captured sequentially with a short time delay in between, the only factor that might influence any difference between LI and RI analysis. \par
Any data that holds identifiable information of an individual is sensitive; more so when it involves children. A two tiered privacy protocol is adapted to protect the identity of the subjects. Due to privacy reasons we do not share any images in this paper unless absolutely needed (falsely rejected image samples and image removed from the database upon manual cleaning) to convey our message. For the images that are included, the iris patterns are obscured. Protecting the privacy of participants in research study is the responsibility of the researchers. We are aware of the lack of children biometric data in the scientific community and the importance of data in advancing research related to child biometrics. The scope of such research has the potential for world-wide applications. We are sharing\cite{CLIC} our dataset through BEAT \cite{anjos2017beat} platform for algorithm testing for research purposes. Protecting the privacy of the data and the individuals, the platform will restrict public access to the data, while giving feedback on the statistics of the performance of the algorithms. In addition to BEAT, we are directly sharing\cite{CLIC}  the Verieye match scores, dilation information and quality scores associated with EA and TF that are used in all of the analyses in this paper.  We are also sharing the data via BEAT and providing VeriEye quality scores associated with each image/pair of images. \par

Further scrutiny on the causes of failure to acquire (FTA) could assist hardware development and application of biometrics for children. FTA is a concern in this study. We noted approximately 25 subjects failed to be captured in one or more sessions. To improve upon the issue, we incorporated an assistive aid from 6th session in the form of a flexible articulating arm for the iris camera mount with a stabilizer handle allowing the collector to adjust the optimal distance required by the camera from the target (refer Figure~\ref{fig:Arm}) and thus essentially eliminating FTA in the 6th session. We also included a second binocular sensor to support the data collection. However, those data are not part of this study. On average, biometrics are presented to the sensor thrice before concluding on FTA. Pushing too hard to collect data from children from whom we failed to acquire, would be harsh, keeping in mind the younger age of the children, which may induce discomfort and possible withdrawal from voluntary participation. Our protocol restricts aggressive data collection from children. The hardware available and used in this study are primarily designed for adults with inbuilt quality assessment setup for enrollment with no manual control. We believe if the systems were designed specifically for children, FTA would be improved.\par
Different studies employ different algorithms to extract features. Commercial algorithms are black-boxes with no public information on the techniques used for image to feature domain transformation, what (if any) methods are adopted to address different quality deformations (like dilation, rotation etc.) or matching techniques. Environmental factors may vary between sessions, like illumination and weather, affecting the captured image. Induced factors like medication, can also affect iris. Thus, the features under investigation are an essential factor in concluding the effect of aging on iris biometric. This reminds us of the need for further research on invariant features for iris recognition. In this study we have investigated the impact of difference in time between enrollment and probe on the iris recognition performance in children under a semi-controlled environment  using a commercial software VeriEye \cite{VerieyeSDK}. 

\section{Conclusion}\label{conclusion}

We conclude by summarizing the answers to the questions, that this work majorly focused on - \par
\textbf{\textit{ Does aging change the iris structure to a point that impacts the use of iris for biometric recognition in children?} }From the point of view of biometric applications, this study indicates the viability of the use of iris as a biometric modality in a 3 year TF on the population with the enrollment age group of 4 to 11 years when the quality of iris image is maintained and FTA are addressed. Statistical analysis demonstrated a minor decay in match score over time i.e. a very small aging effect, much smaller than other typical sources of variation; the impact is practically insignificant. The statistical analysis accounts only for a time-frame of 3 years. The effect on match score is negligible in the 3-year TF of this study. \par

\textbf{\textit{ If growth in children impacts iris recognition performance, is there an age at which these impacts are no longer seen?}} Our study concluded that iris recognition is effective in children for the age group studied i.e between 4 to 11 years. We noted no impact in the iris recognition performance based on age. A positive correlation between enrollment age and match score is indicated in the statistical model. This is expected considering more control and stability with age when presenting the biometric to the system, as well as experience with usage over years. The current study cannot set the minimal age below which iris recognition is ineffective due to three major factors. First, ages below four which were not included in our current dataset needs to be studied. Second, our conclusion excludes consideration of causes of Failure to Acquire (FTA). Thus, FTA needs to be studied. Third, a larger bin size at each age is needed to make any conclusions about a specific age group.\par

Finally, our conclusions are based on research bound by age between 4 yrs to 14 years over a span of three years, and does not translate to life-time conclusions. Some answers are limited by the availability of required research data necessary for analysis. We plan to continue appending our dataset and report the analysis in future towards the unaddressed queries.

 \section*{Acknowledgement}
We extend our gratitude to the Potsdam Elementary and Middle School administration, staff, students and the parents of the participants for supporting our research and the greater goal of scientific contribution to society. This project would not have been successful without the efforts and the valuable time put in by all data collectors including David Yambay, Ganghee Jang, Keivan Bahmani, Sandip Purnapatra and Richard Plesh. This work is financially supported by Center for Identification Technology Research (CITeR) and National Science Foundation (NSF)(\# 1650503).

\bibliographystyle{IEEEtran}
\bibliography{bibliography}
\vspace{-1.25cm}
\vskip 0pt plus -1fil
\begin{IEEEbiography}[{\includegraphics[width=1in,height=1.25in,clip,keepaspectratio]{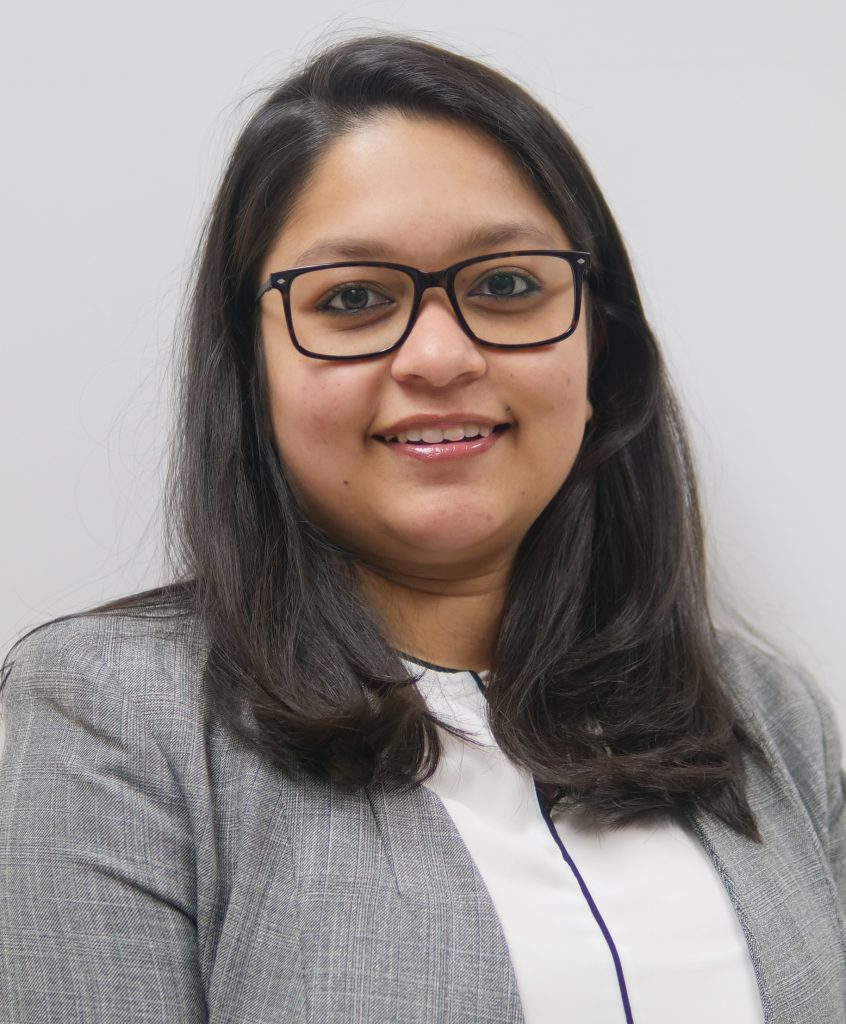}}]{Priyanka Das}
Priyanka Das is a research assistant at Clarkson University pursuing her Ph.D in the Department of Electrical and Computer Engineering, with focus in iris biometrics. She completed her ME in Biomedical Engineering from Jadavpur University, India in 2016 and her Bachelor's in Biomedical Engineering in 2014. 
\end{IEEEbiography}
\vspace{-1 cm}
\vskip 0pt plus -1fil
\begin{IEEEbiography}[{\includegraphics[width=1in,height=1.25in,clip,keepaspectratio]{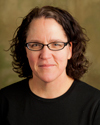}}]{Laura Holsopple}
Laura Holsopple is the Managing Director of the Center for Identification Technology Research (CITeR), an NSF funded Industry University Cooperative Research Center (IUCRC) focusing on the science of biometrics.  Drawing on her 15+ years from Industry Manufacturing and Development, Laura has worked with CITeR over the last 10 years to promote the Center's research and bring continual improvement to the CITeR community and the affiliate members served by membership in the center.
\end{IEEEbiography}
\vspace{-1 cm}
\vskip 0pt plus -1fil
\begin{IEEEbiography}[{\includegraphics[width=1in,height=1.25in,clip,keepaspectratio]{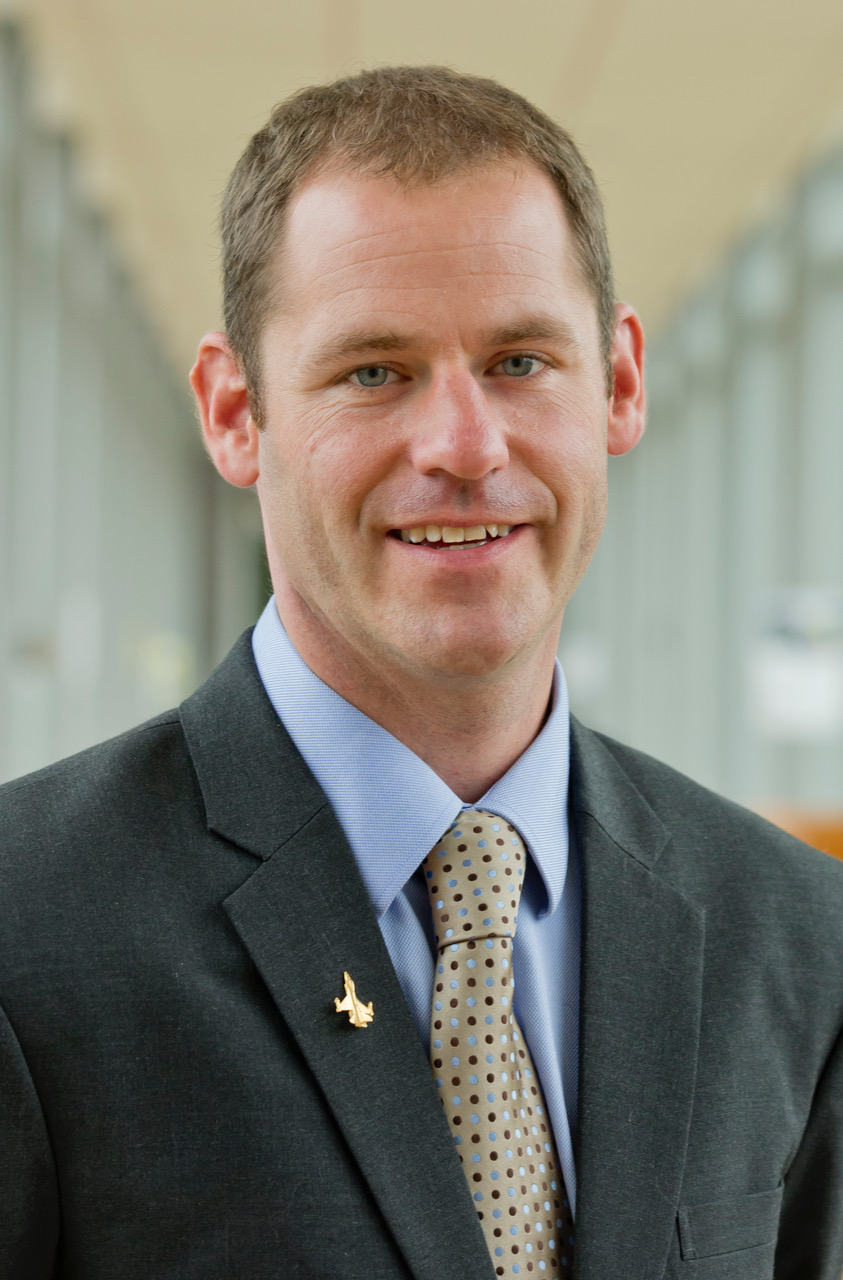}}]{Dan Rissacher} Dr. Dan Rissacher was an Assistant Research Professor at Clarkson University when he founded the study for Biometric Aging in Children. Currently he is an independent researcher at his company, CRIA Corp., a pilot for American Airlines and a Cyber Warfare Operator for the Vermont Air National Guard. Dr. Rissacher earned his  M.S. at Georgia Institute of Technology and Ph.D. at Clarkson with a dissertation demonstrating automated detection of human pain using Neural Networks on EEG data. Since then his research has focused on pattern recognition, data analytics and machine learning.
\end{IEEEbiography}
\vspace{-1 cm}
\vskip 0pt plus -1fil
\begin{IEEEbiography}[{\includegraphics[width=1in,height=1.25in,clip,keepaspectratio]{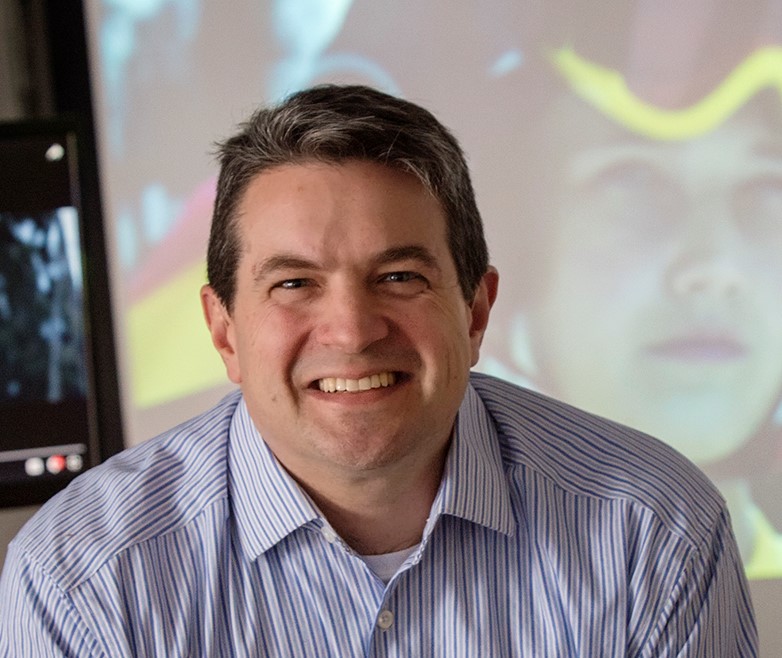}}]{Michael Schuckers}
 Dr. Michael Schuckers is the Charles A. Dana Professor of Statistics at St. Lawrence University in Canton, NY where he also serves as the Director of the Peterson Quantitative Resource
Center. After receiving his doctorate in Statistics from Iowa State University, his academic work has
focused on developing statistical methodology for biometric authentication and applications in sports,
particularly ice hockey. He has also consulted for professional teams in Major League Baseball and the
National Hockey League among other professional sports organizations in his role as co-Founder of
Statistical Sports Consulting, LLC .
\end{IEEEbiography}
\vspace{-1 cm}
\vskip 0pt plus -1fil
\begin{IEEEbiography}[{\includegraphics[width=1in,height=1.25in,clip,keepaspectratio]{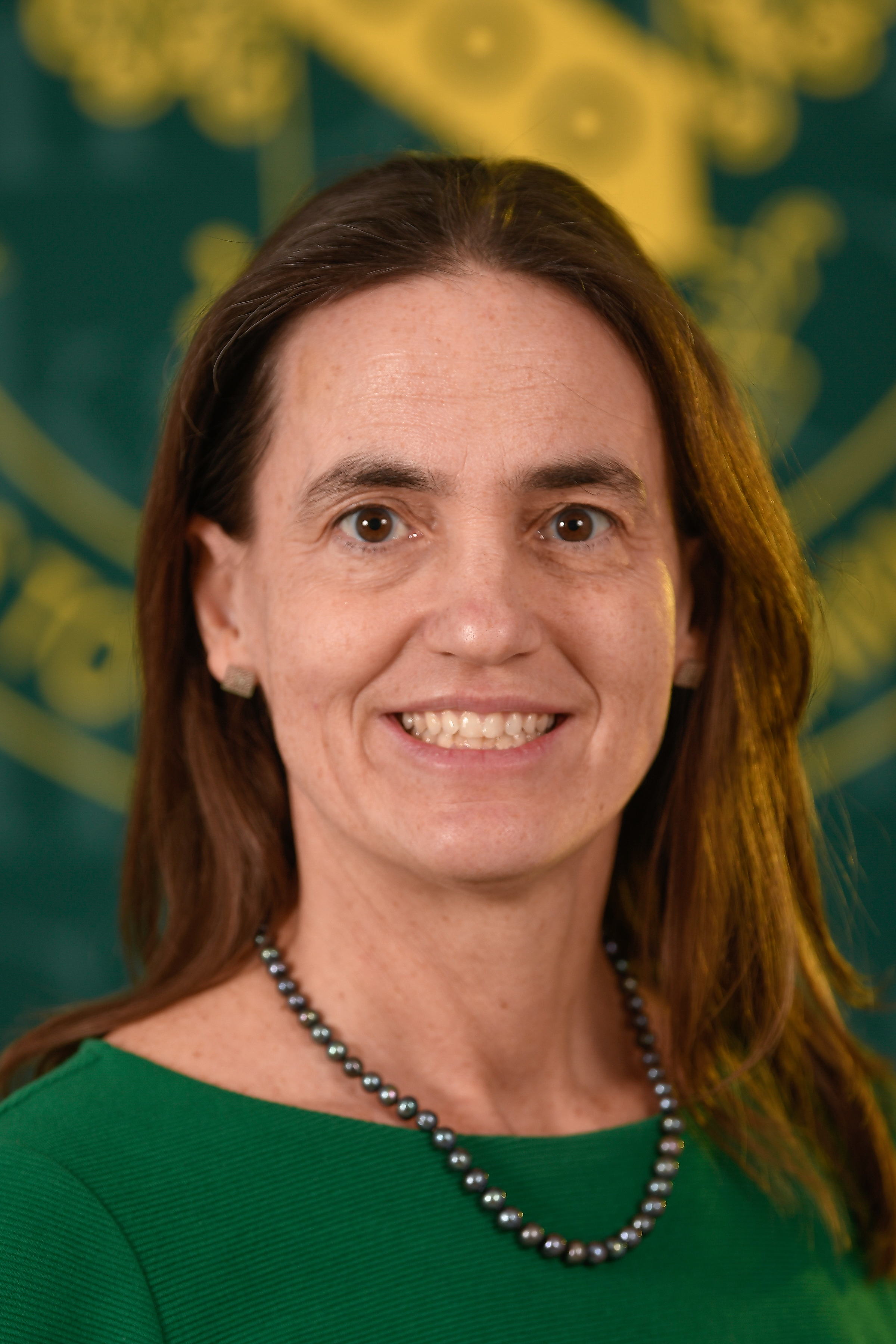}}]{Stephanie Schuckers}
Dr. Stephanie Schuckers is the Paynter-Krigman Endowed Professor in Engineering Science in the Department of Electrical and Computer Engineering at Clarkson University and serves as the Director of the Center for Identification Technology Research (CITeR), a National Science Foundation Industry/University Cooperative Research Center. She received her doctoral degree in Electrical Engineering from The University of Michigan. Professor Schuckers research focuses on processing and interpreting signals which arise from the human body. Her work is funded from various sources, including National Science Foundation, Department of Homeland Security, and private industry, among others.  She has started her own business, testified for the US Congress, and has over 100 other academic publications.
\end{IEEEbiography}

\end{document}